\documentclass[twocolumn,secnumarabic,amssymb, nobibnotes, aps, pra,showpacs,preprintnumbers]{revtex4-1}
\usepackage{amssymb}
\usepackage{array}
\usepackage{graphicx}
\usepackage{ifpdf}

\def\beq{\begin{equation}}
\def\eeq{\end{equation}}
\def\bea{\begin{eqnarray}}
\def\eea{\end{eqnarray}}
\def\no{\nonumber}

\begin{document}

\title{ Hessian matrix, specific heats, Nambu brackets,  and thermodynamic geometry}%

\author{Seyed Ali Hosseini Mansoori}
\email{sa.hosseinimansoori@ph.iut.ac.ir}
\author{Behrouz\ Mirza}
\email{b.mirza@cc.iut.ac.ir}

\affiliation{Department\ of\ Physics,\
Isfahan\ University\ of\ Technology,\ Isfahan\ 84156-83111,\ Iran}
\author{Mohamadreza\ Fazel}
\email{fazel@unm.edu}

\affiliation{Department\ of\ Physics,\ University\ of\ New\ Mexico,\ Albuquerque,\ New\ Mexico,\ 87131-0001,\ USA}
\date{\today}%

\begin{abstract}
As an extension to our earlier work \cite{Mirza2}, we employ the Nambu brackets to prove that the divergences of  heat capacities correspond to their counterparts  in thermodynamic geometry. We also  obtain a simple representation for  the conformal transformations that connect different thermodynamics metrics to each other. Using our bracket approach, we obtain  interesting exact relations between the Hessian matrix with any number of parameters and  specific heat capacities.  Finally, we employ this  approach to investigate some thermodynamic properties of the Meyers-Perry black holes with three spins.
\end{abstract}


\maketitle

\section{Introduction}\label{a}
Stephen Hawking \cite{L1} showed that the area, $A$, of the event horizon
of a black hole never decreases in  physical processes. It was later pointed out by Bekenstein \cite{L2} that this outcome is similar to the statement of the ordinary second law of thermodynamics, namely that the total entropy, $S$, of an isolated system never decreases. Consequently, Bekenstein proposed that the entropy of a black hole is proportional to its area. By considering the propagation of fields in a curved background and close to the event horizon, Hawking later discovered  that black holes radiate with the spectrum of an ideal black body with temperature $T$ \cite{qq11}.  Correspondence between thermodynamics of black holes and the known first and second laws of thermodynamics was further investigated in \cite{aman4}. However, the statistical origin of black hole thermodynamics is still an open problem to realize.

In 1979, Ruppeniner \cite{tun2} proposed a novel method to study the thermodynamics of physical systems, such as black holes. This approach makes use of Reimanian geometry to explore the thermodynamical properties of these systems. The Ruppeiner's metric is based on the fluctuation theory of equilibrium states of the physical systems. In fluctuation theory, every thermodynamic equilibrium system can be characterized as an equilibrium manifold and each point on this manifold is an equilibrium state of the system. The physical system could fluctuate between different points on the equilibrium manifold and, hence, the line element of the stable points on the manifold is proportional to the fluctuation probability. This line element is the basic idea for geometric thermodynamics \cite{rup}. Prior to Ruppeiner, it was Weinhold who  had additionally introduced the phase thermodynamic space and had developed a geometric description of the equilibrium space of thermodynamic states. The Weinhold metric is based on  energy representation \cite{tun1}. Subsequently, the metric was also extended to mass and entropy descriptions, respectively, by Weinhold and Ruppeiner and  they were, consequently, able to
 work out the curvature of the equilibrium manifold. This formulation of thermodynamics equips us with the powerful tools needed for dealing with  thermodynamic problems , as they also been employed to study the thermodynamics of black holes \cite{Ferara, Aman, Shen, Mirza, Medved, Rup3, mbr, mbr1, Abra}.


In this paper, we explore new metrics for the thermodynamic geometry of black holes in the representation of electric charge (Q-metric) and angular momentum (J-metric). We write the partial derivatives in terms of Poisson and Nambu brackets, and employ them to represent the conformal transformations. The general conformal transformation between any two metrics are also obtained using the Poisson and Nambu brackets. Using this approach, the Hessian with an arbitrary number of parameters are also worked out. Moreover, we prove that the divergent points of the specific heats correspond exactly to the singularities of the thermodynamic geometry. The brackets notation is also utilized to work out heat capacities, metric elements, and curvatures.  Finally, we investigate the thermodynamics of the Meyers-Perry black hole with three spins.

The outline of this paper is as follows. In Section II, we introduce Q-metric and J-metric and show that they are related to the Ruppenier and Weinhold metrics through conformal transformations. In Section III, we study the relationship between singularities of the scalar curvature in various representations of thermodynamic geometry and phase transitions of different heat capacities. In Section IV, we show that for Meyers-Perry black hole with three spins, the phase transitions of heat capacities are exactly at the same points of the curvature singularities. Finally in Section V, we discuss our results.
\section{Conformal transformation between different metrics and bracket notation}
Weinhold was the first to  introduce a geometric formulation of thermodynamics \cite{tun1}. The second derivatives of the internal energy (mass) with respect to the entropy and the other extensive variables of the system give the Weinhold metric as follows:
\begin{equation}
ds_{W}^{2}=\frac{{{\partial }^{2}}M}{\partial {{X}^{i}}\partial {{X}^{j}}}d{{X}^{i}}d{{X}^{j}}.
\label{N2}
\end{equation}
Ruppeiner later developed another geometric formulation of
thermodynamics and statistical mechanics \cite{tun2} in which the second derivatives of entropy with respect to mass and the other extensive variables of the system give the Ruppeiner metric as follows:
\begin{equation}
ds_{R}^{2}=\frac{{{\partial }^{2}}S}{\partial {{X}^{i}}\partial {{X}^{j}}}d{{X}^{i}}d{{X}^{j}}.
\label{N3}
\end{equation}

The line elements of the Weinhold and Ruppeiner geometries are conformally related \cite{R33}:
\begin{equation}
ds_{R}^{2}=\frac{1}{T}\frac{{{\partial }^{2}}M}{\partial {{X}^{i}}\partial {{X}^{j}}}d{{X}^{i}}d{{X}^{j}}
\label{N4}
\end{equation}
where, $X^{i}$ are entropy and the other extensive variables of the system. In the case of the charged black hole (Reissner-Nordstrom), the first law of thermodynamics is:
\begin{equation}
dM=TdS+\Phi dQ
\label{N5}
\end{equation}
where,  $(T,\Phi )$ and   $(S,Q)$ are the intensive and extensive variables of the system, respectively. The Weinhold metric can be rewritten as:
\begin{equation}\label{N6}
ds_{W}^{2}(S,Q)=dTdS+d\Phi dQ
\end{equation}
where, $T$ and $\phi $  are the explicit functions of $Q$ and $S$.
What is more, the first law can be transformed into:
\begin{equation}\label{N7}
dS=\frac{1}{T}dM-\frac{\Phi }{T}dQ,
\end{equation}
which implies that the Ruppeiner metric can be written as follows:
\begin{equation}\label{N8}
ds_{R}^{2}(M,Q)=d\left( \frac{1}{T} \right)dM-d\left( \frac{\Phi }{T} \right)dQ
\end{equation}
The combination of (\ref{N6}) and (\ref{N8}) gives:
\begin{equation}
ds_{R}^{2}(M,Q)=-\frac{1}{T}ds_{W}^{2}(S,Q)
\label{N9}
\end{equation}
 Equation (\ref{N8}) represents a Riemannian geometry, which contains some information about the thermodynamics of the physical system. Specifically, the Ricci scalar of this geometry contains a certain amount of knowledge and is a measure of interaction. It also plays a key role in figuring out the thermodynamic stability of the system. As an example, the divergences of this variable are linked to the phase transitions of the system, which are the singularities of the specific heats \cite{Mirza2}. Investigation of different types of black holes, such as the charged black hole, has revealed that the divergences of the Ricci scalar of the Ruppenier geometry corresponds to the singularities of $C_{\Phi}$. This argument has been proved in \cite{Mirza2} for black holes with three parameters. A generalization of this proof for the black holes with an arbitrary number of parameters is discussed in the third section of this paper.
In this section, we develop a new geometric formulation of thermodynamics in which the singularities of the Ricci scalar are the phase transitions associated with the heat capacity at a fixed temperature ($C_T$) and we dubbed it Q-metric. This argument will be demonstrated in the third section.

The first law of thermodynamics might be recast as follows:
\begin{equation}
dQ=\frac{1}{\Phi }dM-\frac{T}{\Phi }dS
\label{N10}
\end{equation}
Regarding (\ref{N10}), one might infer that the Q-metric is:
\beq
ds_{Q}^{2}(M,S)=d(\frac{1}{\Phi})dM-d(\frac{T}{\Phi })dS
\label{WW1}
\eeq
Therefore its line element is:
\begin{equation}
ds_{Q}^{2}(M,S)=\frac{{{\partial }^{2}}Q}{\partial {{X}^{i}}\partial {{X}^{j}}}d{{X}^{i}}d{{X}^{j}}
\label{N11}
\end{equation}
where, ${{X}^{i}}$s are mass and entropy. Furthermore, we will prove that the phase transitions of $C_Q$ and $C_S$, respectively, coincide with the singularities of the Ricci scalar of the metrics whose elements can be derived from the second derivatives of certain functions with respect to extensive/intensive variables. It should be noted that the proofs will not be easy to present in the conventional algebra; however, we will introduce a new convenient notation by using poisson and Nambu brackets.

As already mentioned, the Ruppeiner and Weinhold metrics are related to each other by the conformal transformation (\ref{N9}). Besides, Q-metrics are conformally related to the Ruppenier and Weinhold metrics. In order to obtain the conformal transformations, we transform Eq. (\ref{WW1}) to (\ref{N12}) below:
\bea
   ds_{Q}^{2}(M,S)
 = \frac{-1}{{{\Phi }^{2}}}d\Phi dM+\frac{1}{{{\Phi }^{2}}}Td\Phi dS-\frac{1}{\Phi }dTdS\no \\
   =-\frac{1}{\Phi }d{{s}^{2}}_{W}(S,Q)\hspace{3cm}
 \label{N12}
\eea
The conformal transformation that connects the $Q$-metric and the Ruppeiner metric is:
\bea
   d{{s}^{2}}_{Q}(M, S)=d(\frac{1}{\Phi })dM-d(\frac{T}{\Phi })dS\hspace{.7cm}  \\
 =\frac{-1}{{{\Phi }^{2}}}d\Phi dM+\frac{1}{{{\Phi }^{2}}}Td\Phi dS-\frac{1}{\Phi }dTdS \no \\ \no
 =\frac{T}{\Phi }d{{s}^{2}}_{R}(M, Q)\hspace{2cm}
 \label{N13}
\eea
Fig. 1 depicts the conformal transformations that connect different metrics.\\
\begin{figure}[tbp]
\centering
\fbox{\includegraphics[scale=.3]{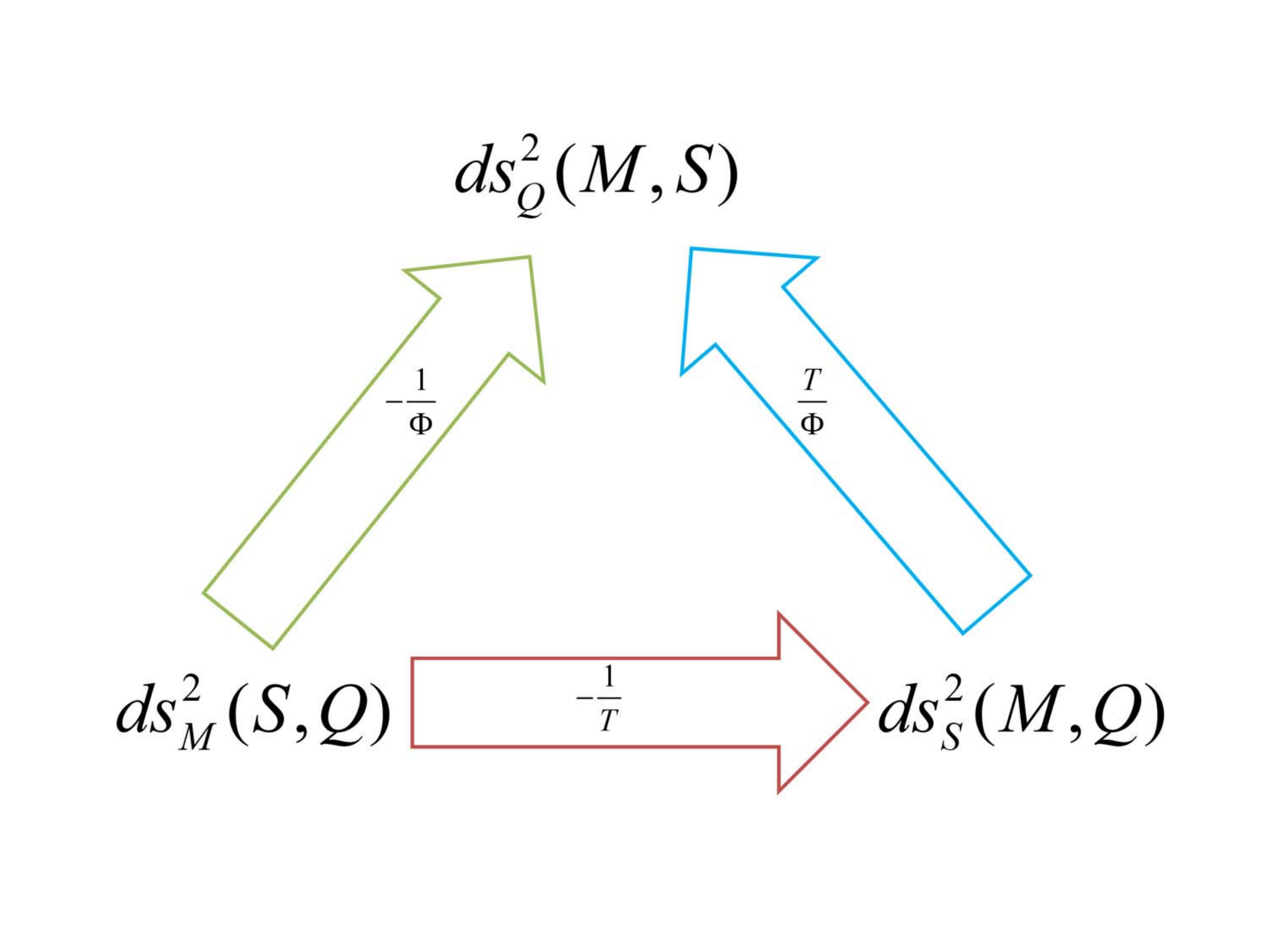}}
\caption{The conformal transformations between various metrics. }
\end{figure}\label{picture1}
It is interesting to have a general formula for the conformal transformations between different geometries. To get such a formula, we write the first law as follows:
\beq
dA=\left(\frac{\partial A}{\partial B}\right)_{C}dB+\left(\frac{\partial A}{\partial C}\right)_{B}dC
\eeq
where, $A$ is a function of $B$ and $C$. Moreover, one can define the A-metric as follows:
\beq
ds_{A}^{2}(B,C)=\frac{{{\partial }^{2}}A}{\partial {{X}^{i}}\partial {{X}^{j}}}d{{X}^{i}}d{{X}^{j}};\,\,\,\,\,{{X}^{i}}=(B,C)
\label{L3}
\eeq
The B-metric and C-metric are also defined as in the following:
\beq
ds_{B}^{2}(A,C)=\frac{{{\partial }^{2}}B}{\partial {{X}^{i}}\partial {{X}^{j}}}d{{X}^{i}}d{{X}^{j}};\,\,\,\,\,{{X}^{i}}=(A,C)
\label{L4}
\eeq

\beq
ds_{C}^{2}(A,B)=\frac{{{\partial }^{2}}C}{\partial {{X}^{i}}\partial {{X}^{j}}}d{{X}^{i}}d{{X}^{j}};\,\,\,\,\,{{X}^{i}}=(A,B)
\label{L5}
\eeq
Contemplating (\ref{L3}) and (\ref{L4}), we reach the relation below, which connects the A-metric to the B-metric via a conformal transformation.
\begin{equation}
d{{s}^{2}}_{A}(B,C)=-{{\left( \frac{\partial A}{\partial B} \right)}_{C}}d{{s}^{2}}_{B}(A,C)
\label{L1}
\end{equation}
It is also straightforward to gain the conformal transformations between A-metric and C-metric and also between B-metric and C-metric.

In the following, we will write the partial derivatives in  terms of Poisson and Nambu brackets. The bracket method has invaluable benefits. It will help us not only to gain a better insight into the different thermodynamic relations but it is also useful for generalizing the thermodynamic formula to arbitrary dimensional phase spaces as we will see in some parts of this paper. Moreover, it allows us to obtain some exact results for the Hessian metric and to calculate heat capacities very easily.

In order to obtain the relation between partial derivatives and Poisson bracket, we should consider $f$,  $g$ and $h$ as explicit functions of ($a, b$). The following useful relation can then be obtained by a simple calculation:

\beq
{{\left( \frac{\partial f}{\partial g} \right)}_{h}}=\frac{{{\left\{ f,h \right\}}_{a,b}}}{{{\left\{ g,h \right\}}_{a,b}}}
\label{N15}
\eeq
where,
\beq
{{\{f,h\}}_{a,b}}={{\left( \frac{\partial f}{\partial a} \right)}_{b}}{{\left( \frac{\partial h}{\partial b} \right)}_{a}}-{{\left( \frac{\partial f}{\partial b} \right)}_{a}}{{\left( \frac{\partial h}{\partial a} \right)}_{b}}
\eeq
is the Poisson bracket.
Consider the following useful identity (for a proof, see Appendix A):\\

\beq\label{y1}
   \left\{ f,g \right\}_{a,b}\left\{ h,k \right\}_{a,b}=\left\{ f,h \right\}_{a,b}\left\{ g,k \right\}_{a,b}-\left\{ f,k \right\}_{a,b}\left\{ g,h \right\}_{a,b}
  \eeq

\noindent It is also simple to demonstrate that:
\begin{equation}
{{\left\{ f,g \right\}}_{a,b}}={{\left\{ f,g \right\}}_{c,d}}{{\left\{ c,d \right\}}_{a,b}},
\label{N19}
\end{equation}
where, $a=a(c,d)$ and $b=b(c,d)$. Therefore, Eq. (\ref{N15}) can be rewritten as:
\begin{equation}
{{\left( \frac{\partial f}{\partial g} \right)}_{h}}=\frac{{{\left\{ f,h \right\}}_{a,b}}}{{{\left\{ g,h \right\}}_{a,b}}}=\frac{{{\left\{ f,h \right\}}_{c,d}}}{{{\left\{ g,h \right\}}_{c,d}}}
\label{N20}
\end{equation}

\noindent The Jacobian transformation in this  notation can be written as:
\begin{equation}
\frac{\partial \left( f,g \right)}{\partial (h,k)}=\left| \begin{array}{cc}
   {{\left( \frac{\partial f}{\partial h} \right)}_{k}} & {{\left( \frac{\partial f}{\partial k} \right)}_{h}}  \\
   {{\left( \frac{\partial g}{\partial h} \right)}_{k}} & {{\left( \frac{\partial g}{\partial k} \right)}_{h}}  \\
\end{array} \right|={{\left\{ f,g \right\}}_{h,k}=\frac{{{\left\{ f,g \right\}}_{a,b}}}{{{\left\{ h,k \right\}}_{a,b}}}}
\label{N17}
\end{equation}

\noindent Now, we utilize (\ref{N20}) to write Eq. (\ref{L1}) as:
\begin{equation}
d{{s}^{2}}_{A}=-{{\left( \frac{\partial A}{\partial B} \right)}_{C}}d{{s}^{2}}_{B}=-\frac{{{\left\{ A,C \right\}}_{a,b}}}{{{\{B,C\}}_{a,b}}}d{{s}^{2}}_{B}
\label{N14}
\end{equation}
where, $B=B(a,b)$ and $C=C(a,b)$. As we will see below, writing the conformal transformation in the bracket representation will significantly ease the task of generalizing it to three or more dimensional spaces.

Contemplating $A=A(B,C,D)$,  where $B(a,b,c)$, $C(a,b,c)$ and $D(a,b,c)$, and following an approach like the two dimensional case, we also show that the conformal transformation, which connects the two metrics in the $A$ and $B$ descriptions, as follows:
\begin{equation}
d{{s}^{2}}_{A}=-\frac{{{\{A,C,D\}}_{a,b,c}}}{{{\{B,C,D\}}_{a,b,c}}}d{{s}^{2}}_{B}
\label{N29}
\end{equation}
in which we have utilized another new notation called the Nambu bracket. For a better understanding, consider the functions $f({{q}_{1}},{{q}_{2}},{{q}_{3}})$, $g({{q}_{1}},{{q}_{2}},{{q}_{3}})$, $h({{q}_{1}},{{q}_{2}},{{q}_{3}})$ and $k({{q}_{1}},{{q}_{2}},{{q}_{3}})$. Now, we want to work out the derivative of $f$ with respect to $g$ when $h$ and $k$ are fixed. Following the above method and doing some simple calculations, we get:
\begin{equation}
{{\left( \frac{\partial f}{\partial g} \right)}_{h,k}}=\frac{{{\left\{ f,h,k \right\}}_{{{q}_{1}},{{q}_{2}},{{q}_{3}}}}}{{{\left\{ g,h,k \right\}}_{{{q}_{1}},{{q}_{2}},{{q}_{3}}}}}
\label{N30}
\end{equation}
where, the Nambu bracket is defined as follows:
\begin{equation}
{{\left\{ f,h,k \right\}}_{{{q}_{1}},{{q}_{2}},{{q}_{3}}}}=\sum\limits_{ijk=1}^{3}{{{\varepsilon }_{ijk}}\frac{\partial f}{\partial {{q}_{i}}}}\frac{\partial h}{\partial {{q}_{j}}}\frac{\partial k}{\partial {{q}_{k}}}
\label{N31}
\end{equation}
and $\varepsilon_{ijk} $ is the Levi-Civita symbol.
Like the Poisson bracket, the Nambu one also has some useful properties such as the
Skew-symmetry:
\begin{equation}
\left\{ {{f}_{1}},{{f}_{2}},{{f}_{3}} \right\}={{(-1)}^{\pi (p)}}\left\{ {{f}_{p(1)}},{{f}_{p(2)}},{{f}_{p(3)}} \right\}
\end{equation}
where, $\pi$ and $p$ are the permutation operator of the subscripts and the parity operator of these permutations, respectively \cite{Nambu1, Nambu2}.\\
The Leibniz rule is expressed by:
\begin{equation}
\left\{ {{f}_{1}}{{f}_{2}},{{f}_{3}},{{f}_{4}} \right\}={{f}_{1}}\left\{ {{f}_{2}},{{f}_{3}},{{f}_{4}} \right\}+\left\{ {{f}_{1}},{{f}_{3}},{{f}_{4}} \right\}{{f}_{2}}
\end{equation}
an the fundamental identity by:
\begin{eqnarray}
\left\{ \left\{ {{f}_{1}},{{f}_{2}},{{f}_{3}} \right\},{{f}_{4}},{{f}_{5}} \right\}+\left\{ {{f}_{3}},\left\{ {{f}_{1}},{{f}_{2}},{{f}_{4}} \right\},{{f}_{5}} \right\}+\\ \no \left\{ {{f}_{3}},{{f}_{4}},\left\{ {{f}_{1}},{{f}_{2}},{{f}_{5}} \right\} \right\}=\left\{ {{f}_{1}},{{f}_{2}},\left\{ {{f}_{3}},{{f}_{4}},{{f}_{5}} \right\} \right\}\hspace{.5cm}
\end{eqnarray}
The Jacobian transformation will be as follows:
\begin{equation}
\frac{\partial \left( {{f}_{1}},{{f}_{2}},{{f}_{3}} \right)}{\partial ({{g}_{1}},{{g}_{2}},{{g}_{3}})}={{\left\{ {{f}_{1}},{{f}_{2}},{{f}_{3}} \right\}}_{{{g}_{1}},{{g}_{2}},{{g}_{3}}}}
\end{equation}
In the above equation, the Nambu bracket is the $3\times 3$ determinant whose elements are the derivative of  ${{f}_{i}}$ with respect to  ${{g}_{i}}$. Considering all the functions as explicit functions of ${{q}_{1}}$, ${{q}_{2}}$ and ${{q}_{3}}$, we have:
\begin{eqnarray}
\frac{\partial \left( {{f}_{1}},{{f}_{2}},{{f}_{3}} \right)}{\partial ({{g}_{1}},{{g}_{2}},{{g}_{3}})}={{\left\{ {{f}_{1}},{{f}_{2}},{{f}_{3}} \right\}}_{{{g}_{1}},{{g}_{2}},{{g}_{3}}}}\\ \no =\frac{{{\left\{ {{f}_{1}},{{f}_{2}},{{f}_{3}} \right\}}_{{{q}_{1}},{{q}_{2}},{{q}_{3}}}}}{{{\left\{ {{g}_{1}},{{g}_{2}},{{g}_{3}} \right\}}_{{{q}_{1}},{{q}_{2}},{{q}_{3}}}}}
\label{N37}
\end{eqnarray}
The conformal transformation between various representations of the thermodynamic metrics of Kerr Newman black hole with the mass, $M(S,Q,J)$,  is depicted schematically in Figure 2.
\begin{figure}[tbp]
\centering
\fbox{\includegraphics[scale=.3]{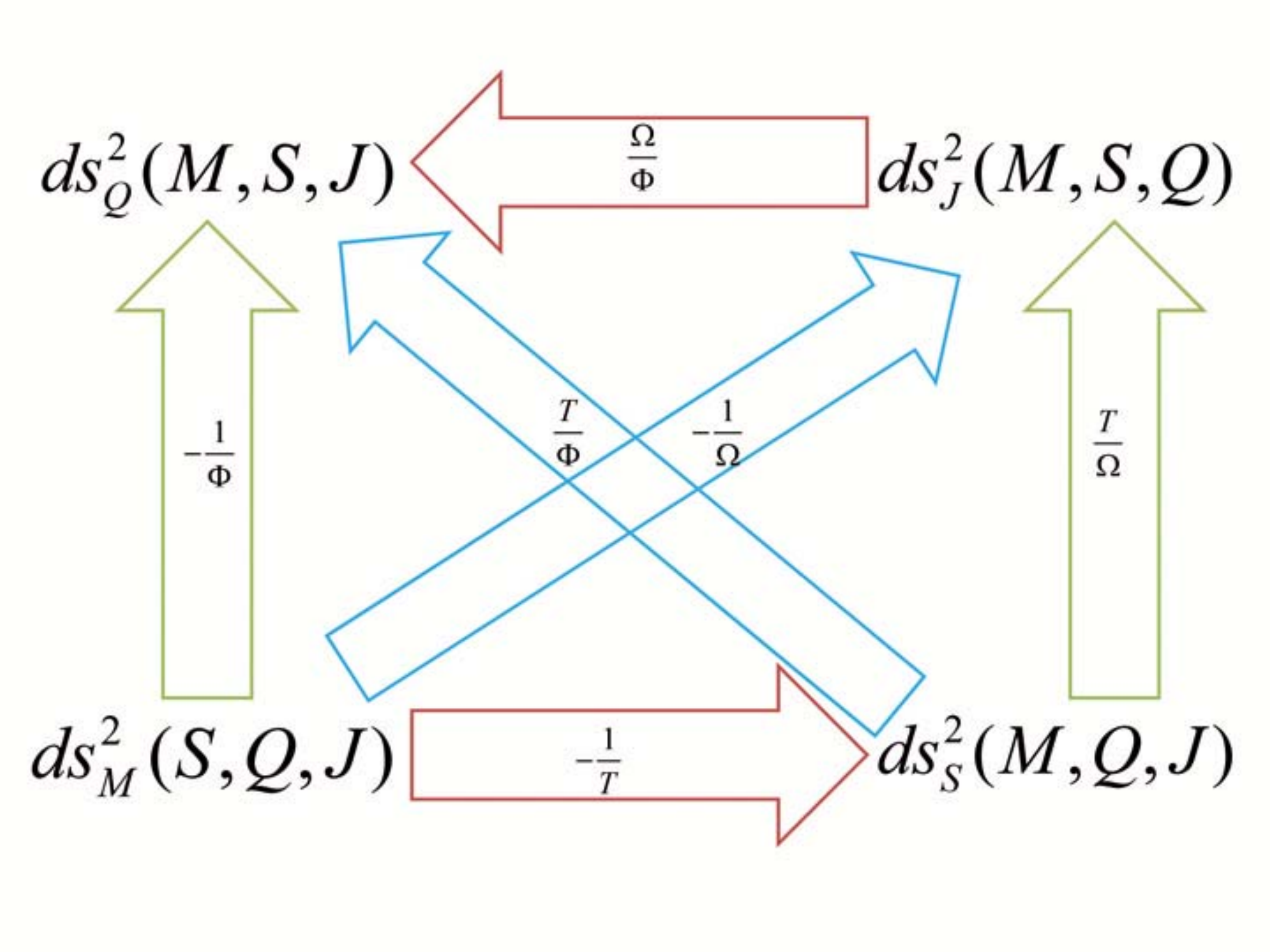} }
\caption{Representation of conformal factor between various metrics for Kerr Newman black hole. Here, $\Omega$ and $\Phi$ are angular velocity and electric potential, respectively.}
\label{picture2}
\end{figure}

Considering $A=A({{X}^{1}},{{X}^{2}},...,{{X}^{n}})$ and ${{X}^{i}}={{X}^{i}}({{x}^{1}},...,{{x}^{n}})$, we show that the conformal transformation that connects the line elements in the A-representation ${{X}^{1}}$-representation is given by:
\begin{equation}
d{{s}^{2}}_{A}=-\frac{{{\{A,{{X}^{2}},...,{{X}^{n}}\}}_{{{x}^{1}},..,{{x}^{n}}}}}{{{\{{{X}^{1}},{{X}^{2}},{{X}^{3}},...,{{X}^{n}}\}}_{{{x}^{1}},...,{{x}^{n}}}}}d{{s}^{2}}_{{{X}^{1}}}
\label{N38}
\end{equation}
In the above equation, we use a new notation as in previous cases and generalize the partial derivative for a function which is a function of $n+1$ variables. Contemplating  $f$, $g$ and ${{h}_{n}}$ ($n=1,\,2,\,3,\,...$) which are functions of ${{q}_{i}},i=1,...,n+1$ variables, we have:
\begin{equation}
{{\left( \frac{\partial f}{\partial g} \right)}_{{{h}_{1}},.....,{{h}_{n}}}}=\frac{{{\left\{ f,{{h}_{1}},...,{{h}_{n}} \right\}}_{{{q}_{1}},{{q}_{2}},...,{{q}_{n+1}}}}}{{{\left\{ g,{{h}_{1}},...,{{h}_{n}} \right\}}_{{{q}_{1}},{{q}_{2}},...,{{q}_{n+1}}}}}
\label{N39}
\end{equation}
where,
\begin{eqnarray}
{{\left\{ f,{{h}_{1}},...,{{h}_{n}} \right\}}_{{{q}_{1}},{{q}_{2}},...,{{q}_{n+1}}}}\hspace{.9cm}\\ \no =\sum\limits_{ijk....l=1}^{n+1}{{{\varepsilon }_{ijk...l}}\frac{\partial f}{\partial {{q}_{i}}}}\frac{\partial {{h}_{1}}}{\partial {{q}_{j}}}\frac{\partial {{h}_{2}}}{\partial {{q}_{k}}}...\frac{\partial {{h}_{n}}}{\partial {{q}_{l}}}
\label{N40}
\end{eqnarray}

Thus, in this Section, we showed that the partial derivatives in two and three dimensions, respectively, are related to the Poisson and Nambu brackets. This method could also be extended to cover every higher dimensional phase spaces. Besides, it has a lot of merits and assists us to perform very difficult algebraic operations effortlessly and the thermodynamics formula in this notation could be simply generalized to any arbitrary dimensional phase spaces.
\section{ Correspondence between transitions and singularities}
In most black holes, the phase transitions are of the second order type. The most important phase transitions in black hole thermodynamics are associated with the divergences of specific heat with a fixed charge, electric potential, temperature, and entropy. Moreover, by studying various charged and rotating black holes, it turns out that the divergences of specific heats correspond to the curvature singularities of various thermodynamic geometries, for which  we provided a proof in \cite{Mirza2}. Using the bracket method, we generalize that work to an arbitrary number of parameters in this section. By making use of the bracket notation, we particularly generalize our former work to an arbitrary number of dimensions.

We start with a brief review of the Ricci scalar and different specific heats. Taking into account the permutation symmetries of the Reimann tensor's indices, one could show that it has only one independent element in two dimensions. Therefore, the Ricci scalar could be written as:
\begin{equation}
R=\frac{2{{R}_{1212}}}{g}
\label{N42}
\end{equation}
where, $g$ is the determinant of the metric. Consider the following equation for black holes with a two-dimensional thermodynamic space as in (\ref{N43}) below:
\begin{equation}
dM=\sum\limits_{i=1}^{2}{{{b}_{i}}d{{q}_{i}}}
\label{N43}
\end{equation}
where, ${{b}_{i}}=\left( \frac{\partial M}{\partial {{q}_{i}}} \right)$  and ${{q}_{i}}$ are intensive and extensive variables, respectively, and the metrics elements in the energy (mass) representation are defined by:
\begin{equation}
{{g}_{ij}}=\left( \frac{\partial {{b}_{i}}}{\partial {{q}_{j}}} \right)
\label{N44}
\end{equation}
The Ricci scalar may be written as follows:
\begin{equation}
R({{q}_{1}},{{q}_{2}})=\frac{\sum\limits_{ijk}{{{\varepsilon }_{ijk}}{{a}_{i}}{{\left\{ {{a}_{j}},{{a}_{k}} \right\}}_{{{q}_{1}},{{q}_{2}}}}}}{-4{{\left( {{\{{{b}_{1}},{{b}_{2}}\}}_{{{q}_{1}},{{q}_{2}}}} \right)}^{2}}}.
\label{h1}
\end{equation}
 where, ${{a}_{1}}={{g}_{11}}$,  ${{a}_{2}}={{g}_{22}}$ and ${{a}_{3}}={{g}_{12}}$.
In general, the phase transitions of the physical systems are attributed to the divergences of the heat capacities, which are one of the most important thermodynamics variables \cite{R53} that show the variation of an extensive variable due to the change in its conjugate parameter.
By defining specific heat at fixed electric potential as following equation:
\bea\label{h5}
{{C}_{\Phi }}=T{{\left( \frac{\partial S}{\partial T} \right)}_{\Phi }}=\frac{T{{\left\{ S,\Phi  \right\}}_{S,Q}}}{{{\left\{ T,\Phi  \right\}}_{S,Q}}}
\eea
it can be demostrated that the phase transitions of specific heat ${{C}_{\Phi }}$ occur accurately in the singularities of $R(S,Q)$, which is the Ricci scalar of the Ruppenier metric. Here, we use the new notation of this paper to show this fact.
It is straightforward that the phase transitions of ${{C}_{\Phi }}$ are the zeros of ${{\left\{ T,\Phi  \right\}}_{S,Q}}$ (See Appendix C). Moreover, we know that the Ruppenier metric is related to Weinhold metric by a conformal transformation.
\beq\label{h14}
g_{ij}^{R}=\frac{1}{T}\left( \frac{{{\partial }^{2}}M}{\partial {{X}^{i}}\partial {{X}^{j}}} \right);{{X}^{i}}=(S,Q)
\eeq
\begin{table*}
\begin{tabular*}{\textwidth}{@{\extracolsep{\fill}}lrl@{}}
\hline
$ {{g}_{SS}}=\frac{{{M}_{SS}}}{T}=\frac{1}{T}{{\left( \frac{\partial T}{\partial S} \right)}_{Q}}=\frac{1}{T}\frac{{{\left\{ T,Q \right\}}_{S,Q}}}{{{\left\{ S,Q \right\}}_{S,Q}}}=\frac{1}{{{C}_{Q}}}=\frac{{{C}_{T}}}{{{C}_{\Phi }}{{C}_{S}}}$   \\
${{g}_{QQ}}=\frac{{{M}_{QQ}}}{T}=\frac{1}{T}{{\left( \frac{\partial \Phi }{\partial Q} \right)}_{S}}=\frac{1}{T}\frac{{{\left\{ \Phi ,S \right\}}_{S,Q}}}{{{\left\{ Q,S \right\}}_{S,Q}}}=\frac{1}{T}\frac{{{\left\{ \Phi ,S \right\}}_{S,Q}}}{{{\left\{ Q,S \right\}}_{S,Q}}}\frac{{{\left\{ T,Q \right\}}_{S,Q}}}{{{\left\{ T,Q \right\}}_{S,Q}}}\frac{{{\left\{ T,\Phi  \right\}}_{S,Q}}}{{{\left\{ T,\Phi  \right\}}_{S,Q}}}==-\frac{{{C}_{\Phi }}}{TQk{{C}_{Q}}}=\frac{1}{T{{C}_{S}}}$\\
$ {{g}_{SQ}}=\frac{{{M}_{SQ}}}{T}=\frac{1}{T}{{\left( \frac{\partial T}{\partial Q} \right)}_{S}}=\frac{1}{T}\frac{{{\left\{ T,S \right\}}_{S,Q}}}{{{\left\{ Q,S \right\}}_{S,Q}}}=\frac{1}{T}\frac{{{\left\{ Q,\Phi  \right\}}_{S,Q}}}{{{\left\{ Q,S \right\}}_{S,Q}}}\frac{{{\left\{ T,\Phi  \right\}}_{S,Q}}}{{{\left\{ T,\Phi  \right\}}_{S,Q}}}\frac{{{\left\{ T,Q \right\}}_{S,Q}}}{{{\left\{ T,Q \right\}}_{S,Q}}}=\frac{\alpha }{k{{C}_{Q}}}=-\frac{Q\alpha }{{{C}_{\Phi }}{{C}_{S}}}$\\
\hline
\end{tabular*}
\caption{  The elements of the two dimensional Ruppenier metric.}.
\end{table*}\label{table20}
The elements of the Ruppenier metric are listed in Table I.
Considering the scalar Ricci (\ref{h1}), it is easy to show that the denominator is given by:
\begin{eqnarray}
\left| \begin{array}{cc}
   \frac{{{M}_{SS}}}{T} & \frac{{{M}_{SQ}}}{T}  \\
   \frac{{{M}_{SQ}}}{T} & \frac{{{M}_{QQ}}}{T}  \\
\end{array} \right|=\frac{1}{{{T}^{2}}}{{\left\{ T,\Phi  \right\}}_{S,Q}}\no \\=\frac{{{\left\{ T,\Phi  \right\}}_{S,Q}}}{{{T}^{2}}}\frac{{{\left\{ S,\Phi  \right\}}_{S,Q}}}{{{\left\{ S,\Phi  \right\}}_{S,Q}}}\frac{{{\left\{ S,Q \right\}}_{S,Q}}}{{{\left\{ S,Q \right\}}_{S,Q}}}
\label{m1}\no \\
=\frac{1}{{{T}^{2}}}\left( \frac{{{\left\{ T,\Phi  \right\}}_{S,Q}}}{{{\left\{ S,\Phi  \right\}}_{S,Q}}} \right)\left( \frac{{{\left\{ S,\Phi  \right\}}_{S,Q}}}{{{\left\{ S,Q \right\}}_{S,Q}}} \right)\no \\=\frac{1}{{{T}^{2}}}\left( \frac{T}{{{C}_{\Phi }}} \right)\left( \frac{1}{{{C}_{S}}} \right)=\frac{1}{T{{C}_{\Phi }}{{C}_{S}}}
\end{eqnarray}
where ${{\left\{ S,Q \right\}}_{S,Q}}=1$ and ${{C}_{S}}$ is defined as follows:
 \beq\label{h6}
{{C}_{S}}={{\left( \frac{\partial Q}{\partial \Phi } \right)}_{S}}=\frac{{{\left\{ Q,S \right\}}_{S,Q}}}{{{\left\{ \Phi ,S \right\}}_{S,Q}}}=\frac{{{\left\{ Q,S \right\}}_{S,\Phi }}}{{{\left\{ \Phi ,S \right\}}_{S,\Phi }}}
\eeq
Thus, it is easy to see that the singularities of $R(S,Q)$ are the roots of  ${{\left\{ T,\Phi  \right\}}_{S,Q}}$. So, we assert that the phase transitions of ${{C}_{\Phi}}$ correspond precisely to the singularities of $R(S,Q)$.
Moreover, we can rewrite the Ricci scalar $R(S,Q)$ using the metrics elements in Table I as the following equation.
\begin{widetext}
\beq
R=\frac{\frac{{{C}_{T}}}{{{C}_{\Phi }}{{C}_{S}}}{{\left\{ \frac{1}{T{{C}_{S}}},-\frac{Q\alpha }{{{C}_{\Phi }}{{C}_{S}}} \right\}}_{S,Q}}-\frac{1}{T{{C}_{S}}}{{\left\{ -\frac{Q\alpha }{{{C}_{\Phi }}{{C}_{S}}},\frac{{{C}_{T}}}{{{C}_{\Phi }}{{C}_{S}}} \right\}}_{S,Q}}-\frac{Q\alpha }{{{C}_{\Phi }}{{C}_{S}}}{{\left\{ \frac{{{C}_{T}}}{{{C}_{\Phi }}{{C}_{S}}},\frac{1}{T{{C}_{S}}} \right\}}_{S,Q}}}{-2{{\left[ \frac{1}{T{{C}_{\Phi }}{{C}_{S}}} \right]}^{2}}}
\eeq
\end{widetext}
where ${{C}_{T}}$ is given by:
\beq\label{h7}
{{C}_{T}}={{\left( \frac{\partial Q}{\partial \Phi } \right)}_{T}}=\frac{{{\left\{ Q,T \right\}}_{S,Q}}}{{{\left\{ \Phi ,T \right\}}_{S,Q}}}
\eeq
In the case of black holes with three parameters, like the Kerr-Newman black hole, one can show that the singularities of $R(S,Q,J)$ and ${{C}_{\Phi ,\Omega }}$ are the same by following a procedure as the above. In order to prove this, we first study the specific heats and Ruppenier metric and we show finally the correspondence between them. For the Kerr-Newman black hole the compressibilities are given in Table II.
\begin{table*}
\begin{tabular*}{\textwidth}{@{\extracolsep{\fill}}lrrl@{}}
\hline \\
$\textbf{Ruppenier metric elements}$\\
\hline \\
$ {{g}_{SS}}=\frac{1}{T}\left( \frac{{{\partial }^{2}}M}{\partial {{S}^{2}}} \right)=\frac{1}{T}{{\left( \frac{\partial T}{\partial S} \right)}_{J,Q}}=\frac{1}{{{C}_{J,Q}}}$ &  ${{g}_{SQ}}=\frac{1}{T}\left( \frac{{{\partial }^{2}}M}{\partial {{S}} \partial {{Q}}} \right)=\frac{1}{T}{{\left( \frac{\partial T}{\partial Q} \right)}_{S,J}}=-\frac{\alpha }{k{{C}_{J,Q}}}$ \\
${{g}_{SJ}}=\frac{1}{T}\left( \frac{{{\partial }^{2}}M}{\partial {{S}} \partial {{J}}} \right)=\frac{1}{T}{{\left( \frac{\partial T}{\partial J} \right)}_{S,Q}}=-\frac{\beta }{\eta {{C}_{Q,J}}}$ & ${{g}_{JQ}}=\frac{1}{T}\left( \frac{{{\partial }^{2}}M}{\partial {{J}} \partial {{Q}}} \right)=\frac{1}{T}{{\left( \frac{\partial \Phi }{\partial J} \right)}_{S,Q}}=\frac{{{C}_{Q,\Phi }}}{T{{C}_{J,Q}}{{C}_{T,Q}}}$\\
${{g}_{QQ}}=\frac{1}{T}\left( \frac{{{\partial }^{2}}M}{\partial {{Q}^2} } \right)=\frac{1}{T}{{\left( \frac{\partial \Phi }{\partial Q} \right)}_{J,S}}=\frac{{{C}_{\Phi ,J}}}{QkT{{C}_{Q,J}}}$ & ${{g}_{JJ}}=\frac{1}{T}\left( \frac{{{\partial }^{2}}M}{\partial {{J}^2}} \right)=\frac{1}{T}{{\left( \frac{\partial \Omega }{\partial J} \right)}_{S,Q}}=\frac{{{C}_{Q,\Omega }}}{J\eta T{{C}_{J,Q}}}$ \\
\hline\\
$\textbf{ Compressibility parameters}$\\
\hline\\
$\alpha =\frac{1}{Q}{{\left( \frac{\partial Q}{\partial T} \right)}_{\Phi ,J}}=\frac{1}{Q}\frac{{{\left\{ Q,\Phi ,J \right\}}_{S,Q,J}}}{{{\left\{ T,\Phi ,J \right\}}_{S,Q,J}}}$ & $\beta =\frac{1}{J}{{\left( \frac{\partial J}{\partial T} \right)}_{\Omega ,Q}}=\frac{1}{Q}\frac{{{\left\{ J,\Omega ,Q \right\}}_{S,Q,J}}}{{{\left\{ T,\Omega ,Q \right\}}_{S,Q,J}}}$  \\
 $k=-\frac{1}{Q}{{\left( \frac{\partial Q}{\partial \Phi } \right)}_{T,J}}=-\frac{1}{Q}\frac{{{\left\{ Q,T,J \right\}}_{S,Q,J}}}{{{\left\{ \Phi ,T,J \right\}}_{S,Q,J}}}$&$\eta =-\frac{1}{J}{{\left( \frac{\partial J}{\partial \Omega } \right)}_{T,Q}}=-\frac{1}{J}\frac{{{\left\{ J,T,Q \right\}}_{S,Q,J}}}{{{\left\{ \Omega ,T,Q \right\}}_{S,Q,J}}}$ \\
 \hline\\
\end{tabular*}
\caption{  The elements of the Ruppenier metric with three parameters and compressibilities.}
\end{table*}\label{table2}
 Using Maxwell's relation (Appendix B), we can show that:
\beq\label{h18}
{{C}_{\Omega ,J}}={{C}_{\Phi ,Q}}
\eeq
Other interesting relations between heat capacities have been shown in Appendix C.

 The elements of the Ruppenier metric could be defined as in the following equation:
\beq\label{fi1}
g_{ij}^{R}=\frac{1}{T}\left( \frac{{{\partial }^{2}}M}{\partial {{X}^{i}}\partial {{X}^{j}}} \right);{{X}^{i}}=(S,Q,J).
\eeq
These metrics can be written as functions of heat capacities and compressibilities (See Table II).

For the KN black hole, the dominator of the Ricci scalar with three parameters is:
\begin{widetext}
\begin{equation}\label{m2}
\left| \begin{array}{ccc}
   \frac{{{M}_{SS}}}{T} & \frac{{{M}_{SQ}}}{T} & \frac{{{M}_{SJ}}}{T}  \\
   \frac{{{M}_{QS}}}{T} & \frac{{{M}_{QQ}}}{T} & \frac{{{M}_{QJ}}}{T}  \\
   \frac{{{M}_{JS}}}{T} & \frac{{{M}_{JQ}}}{T} & \frac{{{M}_{JJ}}}{T}  \\
\end{array} \right|=\frac{1}{{{T}^{3}}}{{\left\{ T,\Phi ,\Omega  \right\}}_{S,Q,J}} =
\frac{1}{{{T}^{2}}{{C}_{\Phi ,\Omega }}{{C}_{S,\Omega }}{{C}_{S,Q}}}
\hspace{2.15cm}
 \end{equation}
\end{widetext}
Hence, the singularities of the Ricci scalar $R(S,Q,J)$ are the zeros of ${{\left\{ T,\Phi ,\Omega  \right\}}_{S,Q,J}}$. In addition, these zeros are the phase transitions of ${{C}_{\Phi ,\Omega }}$.

  One might also generalize these outcomes to a general case with $(n+1)$ parameters in which the intensive variables are more than three. In this case, the first law of thermodynamics can expressed by the following equation:
   \begin{equation}\label{nn1}
   dM=TdS+\sum\limits_{i=1}^{n}{{{\Phi }_{i}}d{{Q}_{i}}}
   \end{equation}
     where, $T$ and $\Phi_{i}$s are the intensive variables. The metric elements of the thermodynamic geometry can be also defined as:
     \begin{equation}\label{nn2}
     g_{ij}^{R}=\frac{1}{T}\left( \frac{{{\partial }^{2}}M}{\partial {{X}^{i}}\partial {{X}^{j}}} \right);{{X}^{i}}=(S,{{Q}_{1}},{{Q}_{2}},...,{{Q}_{n}})
     \end{equation}
     The scalar curvature $R(S,Q_{1},Q_{2},...,Q_{n})$ is proportional to the inverse of the square determinant of the metric.
\beq\label{r3}
R(S,{{Q}_{1}},{{Q}_{2}},...,{{Q}_{n}})\propto \left| \begin{array}{cccc}
 \noalign{\medskip}   {{g}_{S{{Q}_{1}}}} & {{g}_{S{{Q}_{2}}}} & ... & {{g}_{S{{Q}_{n}}}}  \\
  \noalign{\medskip}  {{g}_{{{Q}_{1}}S}} & {{g}_{{{Q}_{1}}{{Q}_{2}}}} & ... & {{g}_{{{Q}_{1}}{{Q}_{n}}}}  \\
  \noalign{\medskip}  : & : & : & :  \\
  \noalign{\medskip}  {{g}_{{{Q}_{n}}S}} & {{g}_{{{Q}_{n}}{{Q}_{1}}}} & ... & {{g}_{{{Q}_{n}}{{Q}_{n}}}}  \\
\end{array} \right|^{-2}
\eeq
Using Eqs. (\ref{nn1}) and (\ref{nn2}), we can rewrite Eq. (\ref{r3}) as:
\begin{eqnarray}\label{m3}
R(S,{{Q}_{1}},{{Q}_{2}},...,{{Q}_{n}})\propto \hspace{1.5cm} \\ \no{{\left( T^{-(n+1)}{{\left\{ T,{{\Phi }_{1}},{{\Phi }_{2}},...,{{\Phi }_{n}} \right\}}_{S,{{Q}_{1}},{{Q}_{2}},...,{{Q}_{n}}}} \right)}^{-2}}
\end{eqnarray}
On the other hand, the heat capacities can be defined by:
\begin{widetext}
\beq \label{PH1}
{{C}_{{{\Phi }_{1}},{{\Phi }_{2}},...,{{\Phi }_{n}}}}=T{{\left( \frac{\partial S}{\partial T} \right)}_{{{\Phi }_{1}},{{\Phi }_{2}},...,{{\Phi }_{n}}}}=T\frac{{{\left\{ S,{{\Phi }_{1}},{{\Phi }_{2}},...,{{\Phi }_{n}} \right\}}_{S,{{Q}_{1}},...,{{Q}_{n}}}}}{{{\left\{ T,{{\Phi }_{1}},{{\Phi }_{2}},...,{{\Phi }_{n}} \right\}}_{S,{{Q}_{1}},...,{{Q}_{n}}}}}
\eeq
\beq
{{C}_{S,{{\Phi }_{2}},...,{{\Phi }_{n}}}}={{\left( \frac{\partial Q_1}{\partial {\Phi}_{1}} \right)}_{{S},{{\Phi }_{2}},...,{{\Phi }_{n}}}}=\frac{{{\left\{ S,{{Q}_{1}},{{\Phi }_{2}},...,{{\Phi }_{n}} \right\}}_{S,{{Q}_{1}},...,{{Q}_{n}}}}}{{{\left\{ S,{{\Phi }_{1}},{{\Phi }_{2}},...,{{\Phi }_{n}} \right\}}_{S,{{Q}_{1}},...,{{Q}_{n}}}}}
\eeq
\beq
{{C}_{{{Q }_{1}},S,...,{{\Phi }_{n}}}}={{\left( \frac{\partial Q_2}{\partial {\Phi}_{2}} \right)}_{{S},{{Q }_{1}},...,{{\Phi }_{n}}}}=\frac{{{\left\{ S,{{Q}_{1}},{{Q }_{2}},...,{{\Phi }_{n}} \right\}}_{S,{{Q}_{1}},...,{{Q}_{n}}}}}{{{\left\{ S,{{Q }_{1}},{{\Phi }_{2}},...,{{\Phi }_{n}} \right\}}_{S,{{Q}_{1}},...,{{Q}_{n}}}}}
\eeq
\beq
\vdots\,\,\,\,\,\,\,\,\,\,\,\,\,\,\,\,\,\,\,\,\,\,\,\,\,\,\,\,\,\,\,\,\,\,\,\,\,\,\,\,\,\,\,\,\,\,\,\,\,\,\,\,\,\,\,\,\,\,\,\,\,\,\,\,\,\,\,\,\,\ \vdots
\eeq
\beq \label{PH2}
{{C}_{{{Q}_{1}},{{Q }_{2}},...,{S}}}={{\left( \frac{\partial Q_n}{\partial {\Phi}_{n}} \right)}_{{S},{{Q }_{1}},...,{{Q }_{n}}}}=\frac{{{\left\{ S,{{Q}_{1}},{{Q }_{2}},...,{{Q}_{n}} \right\}}_{S,{{Q}_{1}},...,{{Q}_{n}}}}}{{{\left\{ S,{{Q }_{1}},{{Q }_{2}},...,{{\Phi }_{n}} \right\}}_{S,{{Q}_{1}},...,{{Q}_{n}}}}}
\eeq
\end{widetext}
One might get $\left( \begin{array}{c}
   2(n+1)  \\
   n-1  \\
\end{array} \right)$ numbers of relations similar to (\ref{ap1}) between these heat capacities.
To extend our results in (\ref{m1}) and (\ref{m2}) to a general case with an arbitrary number of parameters, we recast (\ref{m3}) to the following form:
\begin{widetext}
\begin{eqnarray}\label{WW5}
\frac{1}{{{T}^{n+1}}}{{\left\{ T,{{\Phi }_{1}},{{\Phi }_{2}},...,{{\Phi }_{n}} \right\}}_{S,{{Q}_{1}},{{Q}_{2}},...,{{Q}_{n}}}}= \hspace{3.5cm}
\\
\frac{1}{{{T}^{n+1}}}{{\left\{ T,{{\Phi }_{1}},{{\Phi }_{2}},...,{{\Phi }_{n}} \right\}}_{S,{{Q}_{1}},{{Q}_{2}},...,{{Q}_{n}}}}\frac{{{\left\{ S,{{\Phi }_{1}},{{\Phi }_{2}},...,{{\Phi }_{n}} \right\}}_{S,{{Q}_{1}},{{Q}_{2}},...,{{Q}_{n}}}}}{{{\left\{ S,{{\Phi }_{1}},{{\Phi }_{2}},...,{{\Phi }_{n}} \right\}}_{S,{{Q}_{1}},{{Q}_{2}},...,{{Q}_{n}}}}}\hspace{1.5cm}
\no \\
\frac{{{\left\{ S,{{Q}_{1}},{{\Phi }_{2}},...,{{\Phi }_{n}} \right\}}_{S,{{Q}_{1}},{{Q}_{2}},...,{{Q}_{n}}}}}{{{\left\{ S,{{Q}_{1}},{{\Phi }_{2}},...,{{\Phi }_{n}} \right\}}_{S,{{Q}_{1}},{{Q}_{2}},...,{{Q}_{n}}}}}\frac{{{\left\{ S,{{Q}_{1}},{{Q}_{2}},...,{{\Phi }_{n}} \right\}}_{S,{{Q}_{1}},{{Q}_{2}},...,{{Q}_{n}}}}}{{{\left\{ S,{{Q}_{1}},{{Q}_{2}},...,{{\Phi }_{n}} \right\}}_{S,{{Q}_{1}},{{Q}_{2}},...,{{Q}_{n}}}}}\hspace{1.9cm}
\no \\
...\frac{{{\left\{ S,{{Q}_{1}},{{Q}_{2}},...,{{Q}_{n}} \right\}}_{S,{{Q}_{1}},{{Q}_{2}},...,{{Q}_{n}}}}}{{{\left\{ S,{{Q}_{1}},{{Q}_{2}},...,{{Q}_{n}} \right\}}_{S,{{Q}_{1}},{{Q}_{2}},...,{{Q}_{n}}}}}=\hspace{3.5cm}
\no \\
\frac{1}{{{T}^{n+1}}}\left( \frac{{{\left\{ T,{{\Phi }_{1}},{{\Phi }_{2}},...,{{\Phi }_{n}} \right\}}_{S,{{Q}_{1}},{{Q}_{2}},...,{{Q}_{n}}}}}{{{\left\{ S,{{\Phi }_{1}},{{\Phi }_{2}},...,{{\Phi }_{n}} \right\}}_{S,{{Q}_{1}},{{Q}_{2}},...,{{Q}_{n}}}}} \right)\left( \frac{{{\left\{ S,{{\Phi }_{1}},{{\Phi }_{2}},...,{{\Phi }_{n}} \right\}}_{S,{{Q}_{1}},{{Q}_{2}},...,{{Q}_{n}}}}}{{{\left\{ S,{{Q}_{1}},{{\Phi }_{2}},...,{{\Phi }_{n}} \right\}}_{S,{{Q}_{1}},{{Q}_{2}},...,{{Q}_{n}}}}} \right)\hspace{.8cm}
\no \\
\left( \frac{{{\left\{ S,{{Q}_{1}},{{\Phi }_{2}},...,{{\Phi }_{n}} \right\}}_{S,{{Q}_{1}},{{Q}_{2}},...,{{Q}_{n}}}}}{{{\left\{ S,{{Q}_{1}},{{Q}_{2}},...,{{\Phi }_{n}} \right\}}_{S,{{Q}_{1}},{{Q}_{2}},...,{{Q}_{n}}}}} \right)...\left( \frac{{{\left\{ S,{{Q}_{1}},{{Q}_{2}},...,{{Q}_{n-1}},{{\Phi }_{n}} \right\}}_{S,{{Q}_{1}},{{Q}_{2}},...,{{Q}_{n}}}}}{{{\left\{ S,{{Q}_{1}},{{Q}_{2}},...,{{Q}_{n-1}},{{Q}_{n}} \right\}}_{S,{{Q}_{1}},{{Q}_{2}},...,{{Q}_{n}}}}} \right)=\hspace{.41cm}
\no \\
\frac{1}{{{T}^{n+1}}}\left( \frac{T}{{{C}_{{{\Phi }_{1}},{{\Phi }_{2}},...,{{\Phi }_{n}}}}} \right)\left( \frac{1}{{{C}_{S,{{\Phi }_{2}},...,{{\Phi }_{n}}}}} \right)\left( \frac{1}{{{C}_{{{Q}_{1}},S,...,{{\Phi }_{n}}}}} \right)...\left( \frac{1}{{{C}_{{{Q}_{1}},{{Q}_{2}},...,{{Q}_{n-1}},S}}} \right)=\hspace{.4cm}
\no \\ \no
\frac{1}{{{T}^{n}}{{C}_{{{\Phi }_{1}},{{\Phi }_{2}},...,{{\Phi }_{n}}}}{{C}_{S,{{\Phi }_{2}},...,{{\Phi }_{n}}}}{{C}_{{{Q}_{1}},S,...,{{\Phi }_{n}}}}...{{C}_{{{Q}_{1}},{{Q}_{2}},...,{{Q}_{n-1}},S}}}.\hspace{2.2cm}
\end{eqnarray}
\end{widetext}
To accomplish (\ref{WW5}), we multiplied, in the first step, the early term by fractions, which are identity, and then the nominator of every fraction by the denominator of the next one, which are the heat capacities (\ref{PH1}-\ref{PH2}), and the nominator of the last fraction equals one, namely  ${{\left\{ S,{{Q }_{1}},{{Q }_{2}},...,{{Q }_{n}} \right\}}_{S,{{Q}_{1}},{{Q}_{2}},...,{{Q}_{n}}}} =1$.
The singularities of the Ricci scalar $R(S,{{Q}_{1}},{{Q}_{2}},...,{{Q}_{n}})$ occur only at the phase transition points of the heat capacity ${{C}_{{{\Phi }_{1}},{{\Phi }_{2}},...,{{\Phi }_{n}}}}$ which are the roots of ${{\left\{ T,{{\Phi }_{1}},{{\Phi }_{2}},...,{{\Phi }_{n}} \right\}}_{S,{{Q}_{1}},{{Q}_{2}},...,{{Q}_{n}}}} $. It is important to note that the multi-term
\begin{eqnarray}
\left({{C}_{S,{{\Phi }_{2}},...,{{\Phi }_{n}}}}{{C}_{{{Q}_{1}},S,...,{{\Phi }_{n}}}}...{{C}_{{{Q}_{1}},{{Q}_{2}},...,{{Q}_{n-1}},S}}\right)^{-1}\\ \no={{\left\{ S,{{\Phi }_{1}},{{\Phi }_{2}},...,{{\Phi }_{n}} \right\}}_{S,{{Q}_{1}},{{Q}_{2}},...,{{Q}_{n}}}}\hspace{1cm}
\end{eqnarray}
has no divergences. What is more, one might also conclude that the Hessian (in $M$ representation) is of the following form:
 \begin{eqnarray}\label{HE1}
 \left| \begin{array}{cccc}
 \noalign{\medskip}   {{M}_{S{{Q}_{1}}}} & {{M}_{S{{Q}_{2}}}} & ... & {{M}_{S{{Q}_{n}}}}  \\
  \noalign{\medskip}  {{M}_{{{Q}_{1}}S}} & {{M}_{{{Q}_{1}}{{Q}_{2}}}} & ... & {{M}_{{{Q}_{1}}{{Q}_{n}}}}  \\
  \noalign{\medskip}  : & : & : & :  \\
  \noalign{\medskip}  {{M}_{{{Q}_{n}}S}} & {{M}_{{{Q}_{n}}{{Q}_{1}}}} & ... & {{M}_{{{Q}_{n}}{{Q}_{n}}}}  \\
\end{array} \right|=\hspace{1.5cm} \\ \no
\frac{T}{{{C}_{{{\Phi }_{1}},{{\Phi }_{2}},...,{{\Phi }_{n}}}}{{C}_{S,{{\Phi }_{2}},...,{{\Phi }_{n}}}}{{C}_{{{Q}_{1}},S,...,{{\Phi }_{n}}}}...{{C}_{{{Q}_{1}},{{Q}_{2}},...,{{Q}_{n-1}},S}}}
 \end{eqnarray}
Clearly,  it would be quite impossible to obtain Equations (\ref{WW5}) and (\ref{HE1}) without the bracket notation used here.
Now, let us to return to the RN black hole and verify the phase transitions of ${{C}_{Q}}$. The interesting point that arises in this case is that the phase transitions of this heat capacity do not correspond to the singularities of $R(S,Q)$; however, they are equal to the divergences of $\overline{R}(S,\Phi )$. The line elements of the metric related to this Ricci scalar are:
\beq\label{h19}
\overline{g}_{ij}^{{}}=\frac{1}{T}\left( \frac{{{\partial }^{2}}\overline{M}}{\partial {{X}^{i}}\partial {{X}^{j}}} \right);{{X}^{i}}=(S,\Phi )
\eeq
where, the enthalpy is defined by:
\beq\label{h20}
\overline{M}(S,\Phi )=M(S,Q(S,\Phi ))-\Phi Q(S,\Phi )
\eeq
Additionally, using Eq. (\ref{N20}), we can rewrite heat capacity as:
\beq\label{h21}
{{C}_{Q}}=T{{\left( \frac{\partial S}{\partial T} \right)}_{Q}}=\frac{T{{\left\{ S,Q \right\}}_{S,Q}}}{{{\left\{ T,Q \right\}}_{S,Q}}}=\frac{T{{\left\{ S,Q \right\}}_{S,\Phi }}}{{{\left\{ T,Q \right\}}_{S,\Phi }}}
\eeq
The phase transitions of ${{C}_{Q}}$ are the roots of the denominator or the roots of ${{\left\{ T,Q \right\}}_{S,\Phi }}$. Making use of (\ref{h1}), the denominator of $\overline{R}(S,\Phi )$ is:
\begin{eqnarray} \label{h22}
\left| \begin{array}{cc}
   \frac{1}{T}\left( \frac{{{\partial }^{2}}\bar{M}}{\partial {{S}^{2}}} \right) & \frac{1}{T}\left( \frac{{{\partial }^{2}}\bar{M}}{\partial S\partial \Phi } \right)  \\
   \frac{1}{T}\left( \frac{{{\partial }^{2}}\bar{M}}{\partial \Phi \partial S} \right) & \frac{1}{T}\left( \frac{{{\partial }^{2}}\bar{M}}{\partial {{\Phi }^{2}}} \right)  \\
\end{array} \right|=-\frac{1}{{{T}^{2}}}{{\left\{ T,Q \right\}}_{S,\Phi }}=-\frac{C_S}{TC_Q}\hspace{2cm}
\end{eqnarray}
It is plain that the divergences of $\overline{R}(S,\Phi )$ are the roots of ${{\left\{ T,Q \right\}}_{S,\Phi }}$. So, we assert that the phase transitions of ${{C}_{Q}}$ are equal to the singularities of $\overline{R}(S,\Phi )$.
Moreover, we should check the correspondances between the phase transitions and the singularities of other heat capacities such as the one at a fixed temperature (${{C}_{T}}\equiv {{\left( \frac{\partial Q}{\partial \Phi } \right)}_{T}}$) or the one with fixed entropy (${{C}_{S}}\equiv {{\left( \frac{\partial Q}{\partial \Phi } \right)}_{S}}$).
One might ask the singularities of which curvature corresponds to the phase transitions of this heat capacity? We argue that the answer is the following $Q$-metric
\beq\label{h24}
g_{ij}^{Q}=-\frac{1}{\Phi }\left( \frac{\partial M}{\partial {{X}^{i}}\partial {{X}^{j}}} \right);{{X}^{i}}=(S,Q).
\eeq
It is clear that both singularities of ${{R}^{Q}}(S,Q)$ and phase transition points of $C_T$ are the roots of ${{\left\{ T,\Phi  \right\}}_{S,Q}}$.
One could show that the phase transitions of $C_Q$ is given by the singularities of the Ricci scalar, $\overline{\overline{R}}(T,Q)$, corresponding to the following metric:
\beq\label{h27}
\overline{\overline{g}}_{ij}^{{}}=-\frac{1}{\Phi }\left( \frac{\partial \overline{\overline{M}}}{\partial {{X}^{i}}\partial {{X}^{j}}} \right);{{X}^{i}}=(Q,T)
\eeq
where, $\overline{\overline{M}}(Q,T)$ is the free Helmholtz energy and is defined as
\beq\label{h28}
\overline{\overline{M}}(Q,T)=M(S(T,Q),Q)-TS(Q,T)
\eeq
 The connection between the phase transitions and the divergences of the Ricci scalar is depicted in Figure 3 for the RN black hole.
 \begin{figure}[h]
\centering
\fbox{\includegraphics[scale=.45]{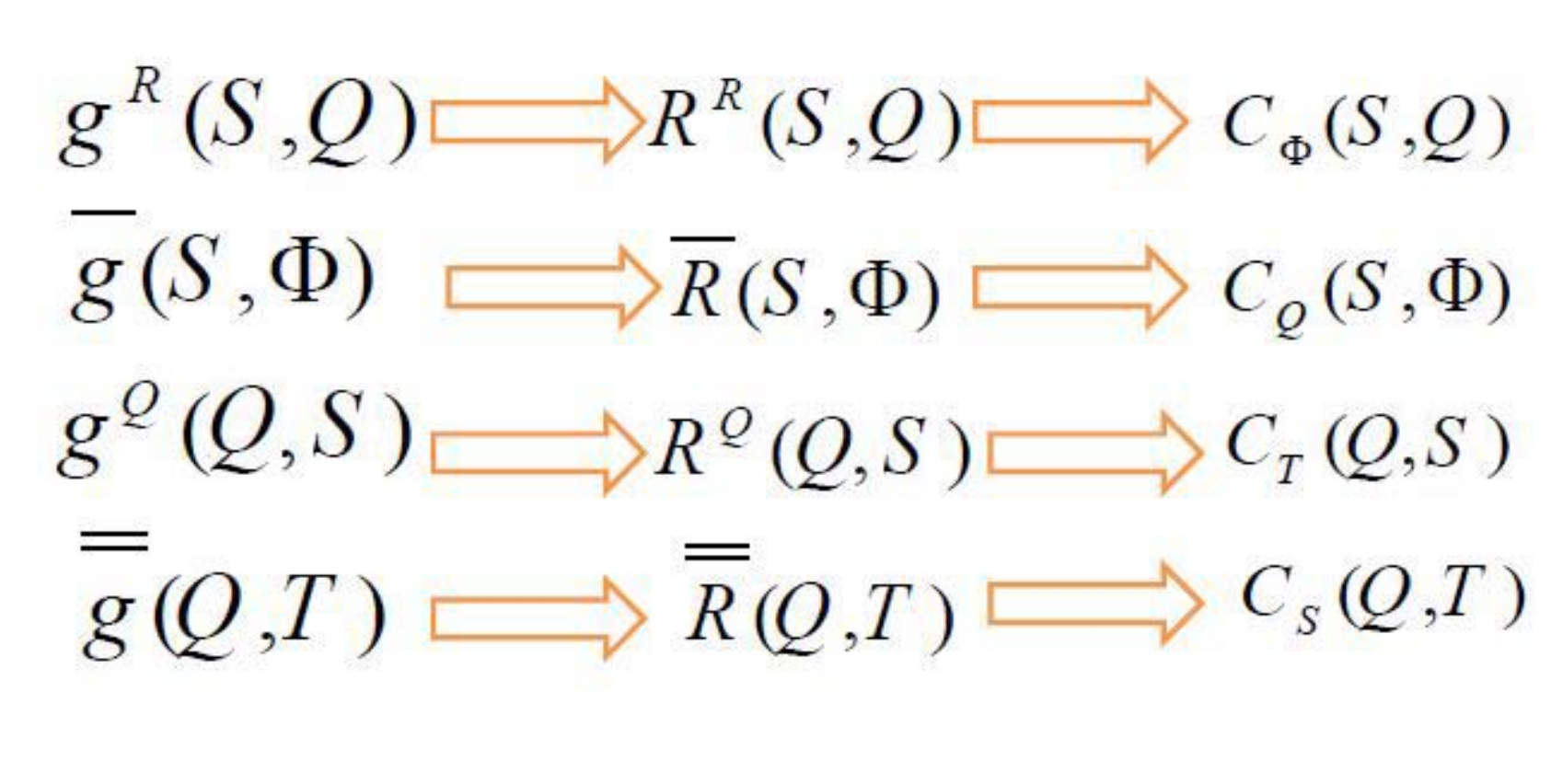} }
\caption{The connection between the singularities of Ricci scalar and phase transitions. }
\end{figure}\label{figure3}

Furthermore, for a black hole with $(n+1)$ parameters, one can define the heat capacity as follows,
\beq
C_{Q_1,Q_2, ..., Q_n}=T\frac{\{S,Q_1 ... Q_n\}_{{\Phi}_1, ...,{\Phi}_n}}{\{T,Q_1, ..., Q_n\}_{{\Phi}_1, ..., {\Phi}_n}}.
\eeq
We will show that the singularities of $R({\Phi}_{1},{\Phi}_{2},...,{\Phi}_{n})$ are the divergences of this heat capacity. Using Eq. (\ref{h19}) and considering the conjugate potential
$\overline{M}=M-{\sum}_{i=1}^{n}{\Phi}_{i}Q_{i}$ one constructs the elements of the metric with the following Ricci scalar:
\begin{eqnarray}
R(S,{{\Phi}_{1}},{{\Phi}_{2}},...,{{\Phi}_{n}})\propto \hspace{2.2cm} \\ \no \left[{\left(\frac{-1}{T}\right)}^n \frac{C_{S,Q_2,...,Q_n}C_{S,{\Phi}_{1},Q_3,...,Q_n}...C_{S,{\Phi}_{1},...,{\Phi}_{n}}}{C_{Q_1,Q_2,...,Q_n}}\right]^{-2}
\end{eqnarray}
Thus, the correspondence between the singularities of the Ricci scalar and the phase transition is clear.

In this section, we proved a correspondence between the heat capacities divergences and the singularities of the Ricci scalar of the thermodynamic geometries. The correct coordinates were also identified in which the Ricci scalar singularities of the metric correspond to the divergences of heat capacity.

\section{The relation between phase transitions and singularities for the Myers-Perry black hole with three spins}

Recently, the Myers-Perry black hole \cite{Myers1} has been of the particular interest due to its rich thermodynamics \cite{Myers2,Myers3,Myers4,Myers5,Myers6,Myers7}. This black hole is a direct generalization of the Kerr black hole in higher dimensions with rotation in more than one plane. The number of the rotating planes $(N)$ are confined by
\beq
N\leq \left[\frac{d-1}{2}\right]
\eeq
where, the bracket gives the integer part and $d$ is the dimensions of the space time. The mass of this black hole in both the odd and even dimensions is given by:
\beq
M=\frac{d-2}{4}{{S}^{\frac{d-3}{d-2}}}{{\prod\limits_{i}^{N}{\left(1+\frac{4J_{i}^{2}}{{{S}^{2}}}\right)}}^{\frac{1}{d-2}}}
\label{MP1}
\eeq
where, $S$ is the entropy and the spin $J_i$ is attributed to the $i$th rotating plane.
 The temperature and angular momentums are given by (\ref{MP2}) and (\ref{MP3}), respectively:
 \beq
 T=\frac{{{S}^{\frac{-1}{d-2}}}}{4}\left[d-3-\sum\limits_{k=1}^{N}{\frac{8J_{k}^{2}}{{{S}^{2}}+4J_{k}^{2}}}\right]
{{\prod\limits_{i}^{N}{\left(1+\frac{4J_{i}^{2}}{{{S}^{2}}}\right)}}^{\frac{1}{d-2}}} \label{MP2}\eeq
 \beq
 {{\Omega }_{p}}=\frac{{{S}^{\frac{d-3}{d-2}}}}{4}\frac{8{{J}_{P}}}{{{S}^{2}}+4J_{P}^{2}}\prod\limits_{i=1}^{N}{\left(1+\frac{4J_{i}^{2}}{{{S}^{2}}}\right)}^{\frac{1}{d-2}}
\label{MP3}
\eeq

 Although the thermodynamics of the Myers-Perry black hole with two rotating planes, equal spins, and a particular number of nonzero equal spins have been extensively explored \cite{Myers2,Myers6,Myers7,Myers8}.The Hawking  temperature for a Myers-Perry black hole with more than two and unequal spins have been obtianed in \cite{Myers8}. However, specific heats and phase transitions have not yet been reported for more than two and unequal spins. This is perhaps due to the complicated algebra that is often associated with the conventional methods for cases with more than two spins. In order to show the power of the bracket notation, we exploit it to derive the thermodynamics of the Myers-Perry black hole with three spinning planes, whereby we gain its phase transitions.

The mass of the Myers-Perry black hole with three spins is given by:
\begin{eqnarray}
M=\frac{\left( d-2 \right)}{4} {S}^{{\frac {d-3}{d-2}}} \left( 1+4\,{\frac{
{J_{{1}}}^{2}}{{S}^{2}}} \right) ^{ \frac{1}{\left( d-2 \right)}}\\ \no \left( 1
+4\,{\frac {{J_{{2}}}^{2}}{{S}^{2}}} \right) ^{ \frac{1}{\left( d-2 \right)}} \left( 1+4\,{\frac{{J_{{3}}}^{2}}{{S}^{2}}} \right) ^{ \frac{1}{\left( d-2
 \right)}}
\end{eqnarray}
We examine the following three heat capacities of Myers-Perry black hole in this work:
\beq
C_{J_1, J_2, J_3}=T{{\left( \frac{\partial S}{\partial T} \right)}_{{{J}_{1}},{{J}_{2}},{{J}_{3}}}}=T\frac{\{S, J_1, J_2, J_3 \}_{S, J_1, J_2, J_3}}{\{T, J_1, J_2, J_3 \}_{S, J_1, J_2, J_3}}
\eeq
\beq
C_{{\Omega}_1, {\Omega}_2, {\Omega}_3}=T{{\left( \frac{\partial S}{\partial T} \right)}_{{{\Omega }_{1}},{{\Omega }_{2}},{{\Omega }_{3}}}}=T\frac{\{S, {\Omega}_1, {\Omega}_2, {\Omega}_3 \}_{S, J_1, J_2, J_3}}{\{T, {\Omega}_1, {\Omega}_2, {\Omega}_3 \}_{S, J_1, J_2, J_3}}
\eeq
\beq
C_{T, {\Omega}_1, {\Omega}_2}={{\left( \frac{\partial {{J}_{3}}}{\partial {{\Omega }_{3}}} \right)}_{T,{{\Omega }_{1}},{{\Omega }_{2}}}}=\frac{\{J_3, T, {\Omega}_1, {\Omega}_2 \}_{S, J_1, J_2, J_3}}{\{{\Omega}_3, T, {\Omega}_1, {\Omega}_2 \}_{S, J_1, J_2, J_3}}
\eeq
The exact statement for these heat capacities have been already derived; however, they appeared to be very long and we would not repeat them here. Instead, we have listed the number of phase transitions in Tables III, IV, and V and some relevant graphs are included, too. Some special cases of phase transitions are presented in Table VI. We observe that the heat capacity $C_{J_1, J_2, J_3}$ has no phase transitions in space time dimensions above 31 ($d=31$), neither has the case $J_1 =J_2$ and $J_3=0$ in dimensions higher than thirteen (d=13). There are either one or two transitions points in the dimensions lower than 31 ($d=31$) for $J_1\neq J_2 \neq J_3$. The heat capacity at fixed angular momentums, $C_{{\Omega}_1, {\Omega}_2, {\Omega}_3}$, has no phase transitions in dimensions below 10 ($d=10$) when $J_1=J_2=J_3$. This heat capacity has either two or three phase transitions when $J_1\neq J_2\neq J_3$. Moreover, the heat capacity $C_{T, {\Omega}_1, {\Omega}_2}$ has not phase transitions when $J_1=J_2$ and $J_3=0$. The Ricci scalar $R(S, {\Omega}_1, {\Omega}_2, {\Omega}_3)$ and $R^J(S, J_1, J_2, J_3)$ are depicted in Figures 12 and 17, respectively. A comparison of these two reveals that their singularities exactly correspond to the phase transitions of $C_{J_1, J_2, J_3}$ and $C_{T, {\Omega}_1, {\Omega}_2}$ in Figures 10 and 16, respectively. Interestingly, we see that the phase transitions in Figure 11 are not the second order phase transitions, rather they might be of higher order types.
\begin{table*}
\begin{tabular*}{\textwidth}{@{\extracolsep{\fill}}lrrrrl@{}}
\hline
$d$ & $J_{1}=J_{2}$ and  $J_{3}=0$ &$J_{1}=J_{2}\neq J_{3}$& $J_{1}=J_{2}=J_{3}$ & $J_{1}\neq J_{2}\neq J_{3}$\\
\hline
$7$& One point &One point& One point& One point\\
\hline
$8$& Two points &One point& One point & One point\\
\hline
$9$& Two points &One point &One point & One point\\
\hline
$10$& Two points &Two points& One point & Two points\\
\hline
$11$& Two points &Two points &Two points & Two points\\
\hline
$12$& Two points & Two points&Two points & Two points\\
\hline
\end{tabular*}
\caption{  The number of phase transitions for $C_{J_{1},J_{2},J_{3}}$. For dimensions above 31, there is no phase transition.}
\end{table*}\label{table2}

\begin{table*}
\begin{tabular*}{\textwidth}{@{\extracolsep{\fill}}lrrrrl@{}}
\hline
$d$ & $J_{1}=J_{2}$ and  $J_{3}=0$ &$J_{1}=J_{2}\neq J_{3}$& $J_{1}=J_{2}=J_{3}$ & $J_{1}\neq J_{2}\neq J_{3}$\\
\hline
$7$& No point &One point& No point& Two point\\
\hline
$8$& One point &One point& No point & Two point\\
\hline
$9$& One point &One point &No point & Two point\\
\hline
$10$& One point &Two points& One point & Three points\\
\hline
$11$& One point &Two points &One point & Three points\\
\hline
$12$& One point & Two points&One point & Three points\\
\hline
\end{tabular*}
\caption{  The number of phase transitions for the heat capacity $C_{\Omega_{1},\Omega_{2},\Omega_{3}}$ in an arbitrary number of dimensions.}
\end{table*}\label{table2}

\begin{table*}
\begin{tabular*}{\textwidth}{@{\extracolsep{\fill}}lrrrrl@{}}
\hline
$d$ & $J_{1}=J_{2}$ and  $J_{3}=0$ &$J_{1}=J_{2}\neq J_{3}$& $J_{1}=J_{2}=J_{3}$ & $J_{1}\neq J_{2}\neq J_{3}$\\
\hline
$7$& No point &One point& One point& Two point\\
\hline
$8$& No point &One point& One point & Two point\\
\hline
$9$& No point &One point &One point & Two point\\
\hline
$10$& No point &Two points& Two points & Three points\\
\hline
$11$& No point &Two points &Two points & two or three points\\
\hline
$12$& No point & Two points&Two points & Three points\\
\hline
\end{tabular*}
\caption{  The number of phase transitions for the heat capacity $C_{T,\Omega_{1},\Omega_{2}}$ with an arbitrary number of dimensions.}
\end{table*}\label{table2}

\begin{table*}
\begin{tabular*}{\textwidth}{@{\extracolsep{\fill}}lrl@{}}
\hline\\
$C_{J_{1}, J_{2}, J_{3}}(J_{1}=J_{2}=J; J_{3}=0)=-{\frac { \left( {S}^{2}+4\,{J}^{2} \right)  \left( d-2 \right)
 \left( d{S}^{2}+4\,d{J}^{2}-3\,{S}^{2}-28\,{J}^{2} \right) S}{-3\,{S}
^{4}-560\,{J}^{4}-8\,d{S}^{2}{J}^{2}-24\,{S}^{2}{J}^{2}+d{S}^{4}+80\,d
{J}^{4}}}$\\
\\
$C_{J_{1}, J_{2}, J_{3}}(J_{1}=J_{2}=J_{3}=J)=-{\frac { \left( {S}^{2}+4\,{J}^{2} \right)  \left( d-2 \right)
 \left( d{S}^{2}+4\,d{J}^{2}-3\,{S}^{2}-36\,{J}^{2} \right) S}{-3\,{S}
^{4}-1008\,{J}^{4}-16\,d{S}^{2}{J}^{2}-24\,{S}^{2}{J}^{2}+d{S}^{4}+112
\,d{J}^{4}}}$\\
\\
$C_{\Omega_{1},\Omega_{2},\Omega_{3}}(J_{1}=J_{2}=J;J_{3}=0)=-{\frac { \left( -24\,{J}^{2}+4\,d{J}^{2}-d{S}^{2}+2\,{S}^{2} \right)
 \left( d{S}^{2}+4\,d{J}^{2}-3\,{S}^{2}-28\,{J}^{2} \right) S}{
 \left( -28\,{J}^{2}+4\,d{J}^{2}+3\,{S}^{2}-d{S}^{2} \right)  \left({
S}^{2}+4\,{J}^{2} \right) }}
$\\
\\
$C_{\Omega_{1},\Omega_{2},\Omega_{3}}(J_{1}=J_{2}=J_{3}=J)=-{\frac { \left( 4\,d{J}^{2}-32\,{J}^{2}+2\,{S}^{2}-d{S}^{2} \right)
 \left( d{S}^{2}+4\,d{J}^{2}-3\,{S}^{2}-36\,{J}^{2} \right) S}{
 \left( 4\,d{J}^{2}-36\,{J}^{2}+3\,{S}^{2}-d{S}^{2} \right)  \left( {S
}^{2}+4\,{J}^{2} \right) }}
$\\
\\
$C_{\Omega_{1},\Omega_{2},\Omega_{3}}(J_{1};J_{2};J_{3}=0)=$\\
$-{\frac { \left(  \left( d-3 \right) {S}^{4}+4\, \left( {J_{{2}}}^{2}+
{J_{{1}}}^{2} \right)  \left( d-5 \right) {S}^{2}+16\,{J_{{1}}}^{2}{J_
{{2}}}^{2} \left( d-7 \right)  \right)  \left(  \left( d-2 \right) {S}
^{4}-4\, \left( {J_{{2}}}^{2}+{J_{{1}}}^{2} \right)  \left( -4+d
 \right) {S}^{2}+16\,{J_{{1}}}^{2}{J_{{2}}}^{2} \left( d-6 \right)
 \right) S}{ \left(  \left( d-3 \right) {S}^{4}-4\, \left( {J_{{2}}}^{
2}+{J_{{1}}}^{2} \right)  \left( d-5 \right) {S}^{2}+16\,{J_{{1}}}^{2}
{J_{{2}}}^{2} \left( d-7 \right)  \right)  \left( {S}^{2}+4\,{J_{{1}}}
^{2} \right)  \left( {S}^{2}+4\,{J_{{2}}}^{2} \right) }}
$\\
\\
$C_{T,\Omega_{1},\Omega_{2}}(J_{1}=J_{2}=J;J_{3}=0)=\frac{{{S}^{\frac{d+3}{d-2}}}}{2}{{\left( {{S}^{2}}+4{{J}^{2}} \right)}^{-\frac{2}{d-2}}}$\\
\\
$C_{T,\Omega_{1},\Omega_{2}}(J_{1};J_{2};J_{3}=0)=\frac{{{S}^{2}}}{2\left( {{S}^{\frac{d-3}{d-2}}} \right)\left( {{\left( \frac{{{S}^{2}}+4{{J}_{1}}^{2}}{{{S}^{2}}} \right)}^{\frac{1}{d-2}}} \right)\left( {{\left( \frac{{{S}^{2}}+4{{J}_{2}}^{2}}{{{S}^{2}}} \right)}^{\frac{1}{d-2}}} \right)}$\\
\\
$C_{T,\Omega_{1},\Omega_{2}}(J_{1}=J_{2}=J_{3}=J)=\frac{-\left( -1728{{J}^{6}}+192{{J}^{6}}d+144{{S}^{2}}{{J}^{4}}-48{{J}^{4}}d{{S}^{2}}-4{{J}^{2}}{{S}^{4}}+4{{J}^{2}}d{{S}^{4}}+3{{S}^{6}}-d{{S}^{6}} \right)}{2\left( 2J-S \right)\left( 2J+S \right)\left( -36{{J}^{2}}+4d{{J}^{2}}+3{{S}^{2}}-d{{S}^{2}} \right){{\left( {{\left( \frac{{{S}^{2}}+4{{J}^{2}}}{{{S}^{2}}} \right)}^{\frac{1}{d-2}}} \right)}^{3}}\left( {{S}^{\frac{d-3}{d-2}}} \right)}$\\
\hline\\
\end{tabular*}
\caption{  The representation of heat capacities in different states.}
\end{table*}\label{table2}

\begin{figure}[tbp]
\centering
\fbox{\includegraphics[scale=.47]{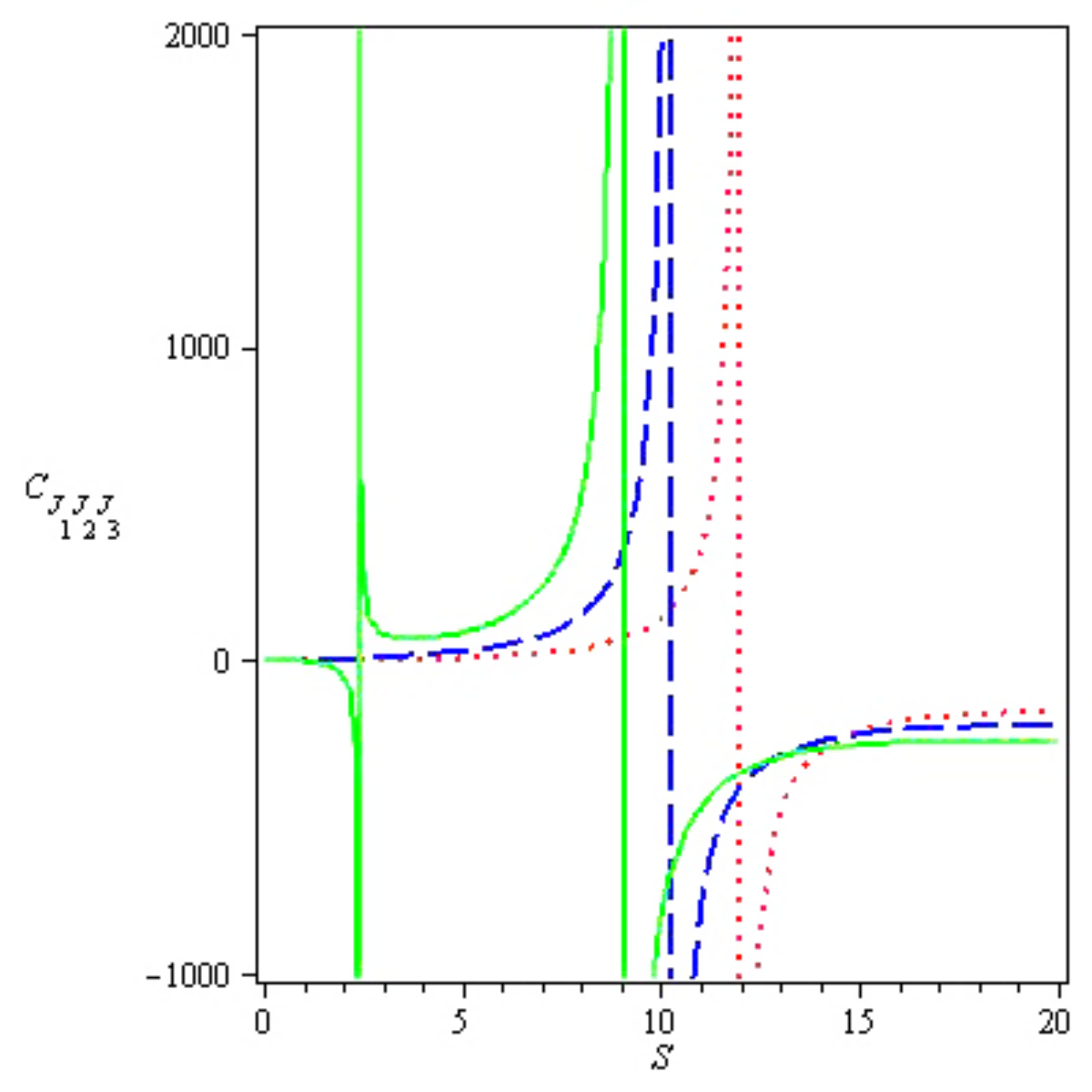}}
\caption{Graph of the the heat capacity $C_{J_{1},J_{2},J_{3}}$ with respect to entropy, $S$, for  $J_{1} $= 1, $J_{2} $= 2 and $J_{3} $= 3. The dot red, dot-dashed blue, and solid green curves correspond to $d=7$, $d=9$ and $d=11$, respectively.   }
\end{figure}
\begin{figure}[tbp]
\centering
\fbox{\includegraphics[scale=.47]{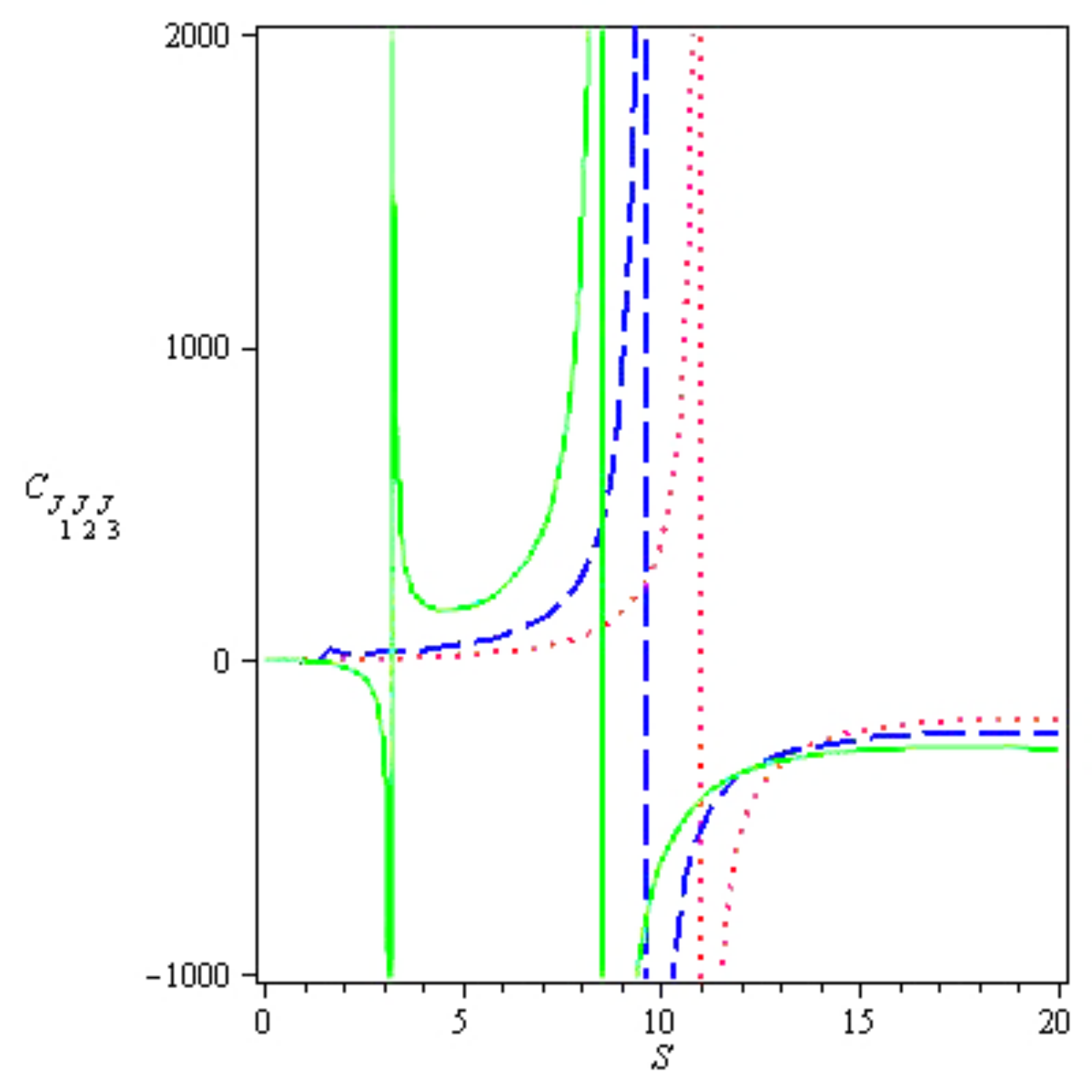}}
\caption{Graph of the the heat capacity $C_{J_{1},J_{2},J_{3}}$ with respect to entropy, $S$, for  $J_{1} $= 1, $J_{2} $= 2 and $J_{3} $= 3. The dot red, dot-dashed blue, and solid green curves correspond to $d=8$, $d=10$ and $d=12$, respectively. }
\end{figure}
\begin{figure}[tbp]
\centering
\fbox{\includegraphics[scale=.47]{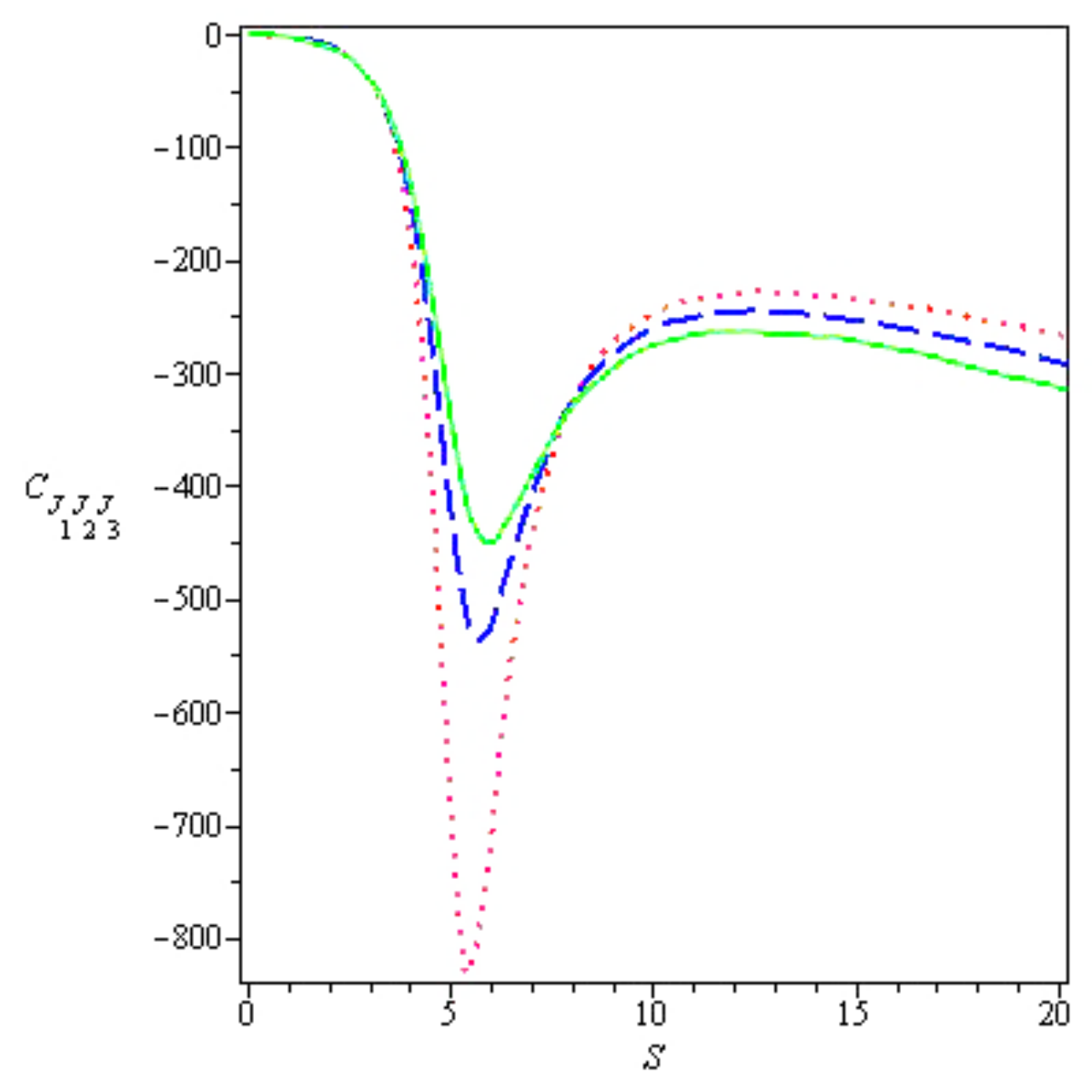}}
\caption{Graph of the the heat capacity $C_{J_{1},J_{2},J_{3}}$ with respect to entropy, $S$, for  $J_{1} = J_{2} $= 2, and $J_{3} $= 0. The dot red, dot-dashed blue, and solid green curves correspond to $d=13$, $d=14$ and $d=15$, respectively. }
\end{figure}
\begin{figure}[tbp]
\centering
\fbox{\includegraphics[scale=.47]{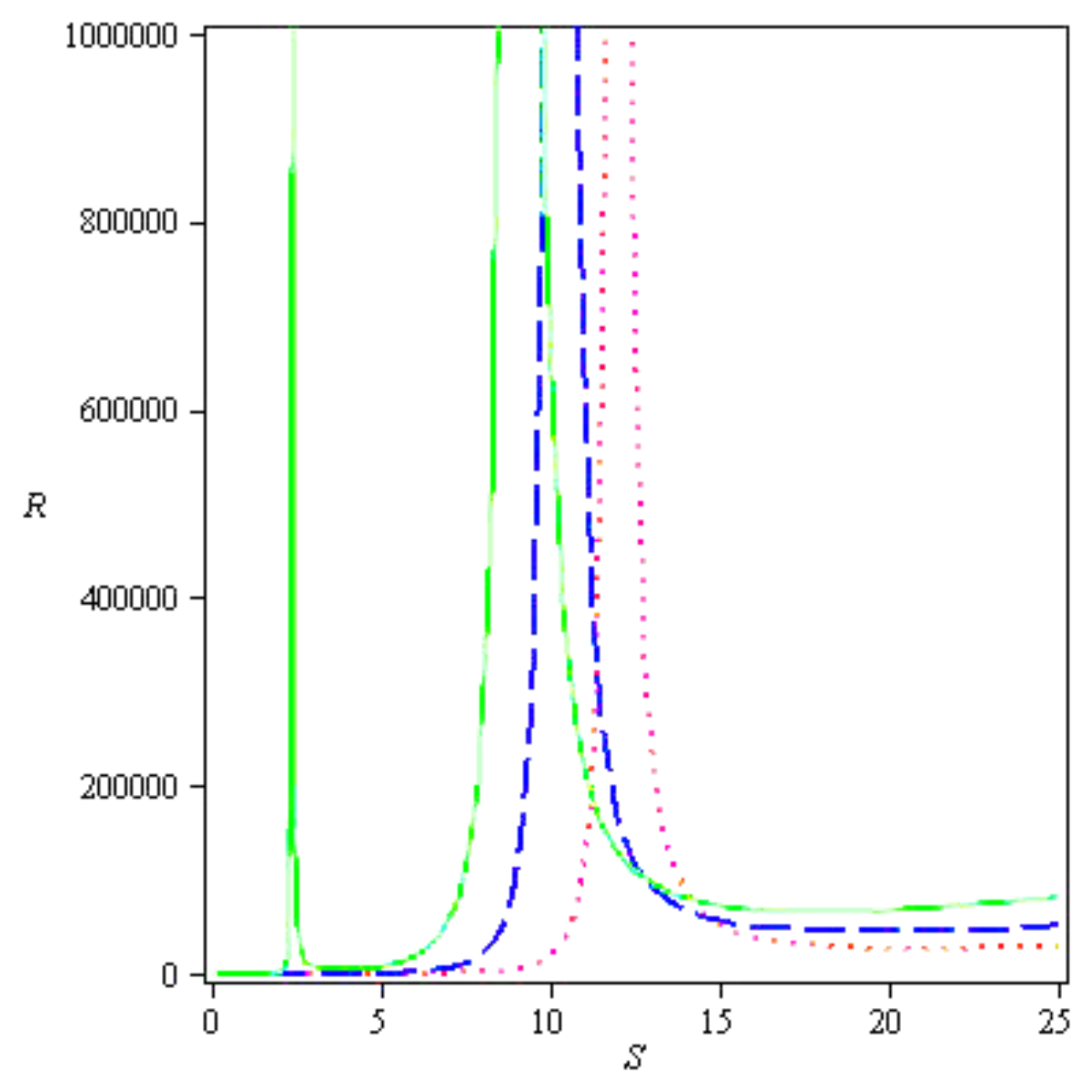}}
\caption{Graph of the the scalar curvature $R(S,\Omega_{1},\Omega_{2},\Omega_{3})$ with respect to entropy, $S$, for  $J_{1} $= 1, $J_{2} $= 2, and $J_{3} $= 3. The dot red, dot-dashed blue, and solid green curves correspond to $d=7$, $d=9$ and $d=11$, respectively. }
\end{figure}
\begin{figure}[tbp]
\centering
\fbox{\includegraphics[scale=.47]{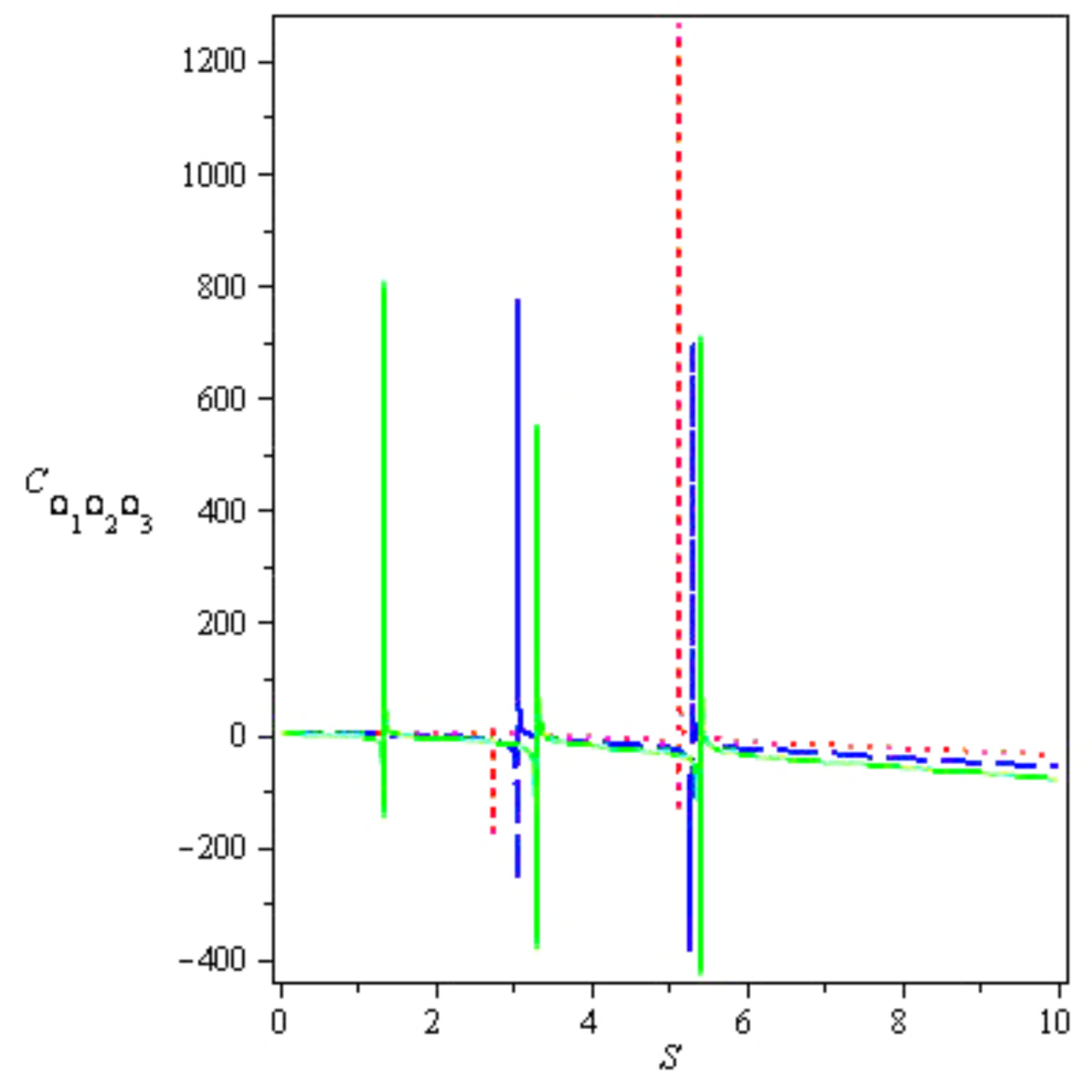} }
\caption{Graph of the the scalar curvature $C_{\Omega_{1},\Omega_{2},\Omega_{3}}$ with respect to entropy, $S$, for  $J_{1} $= 1, $J_{2} $= 2, and $J_{3} $= 3. The dot red, dot-dashed blue, and solid green curves correspond to $d=7$, $d=9$ and $d=11$, respectively. }
\end{figure}
\begin{figure}[tbp]
\centering
\fbox{\includegraphics[scale=.47]{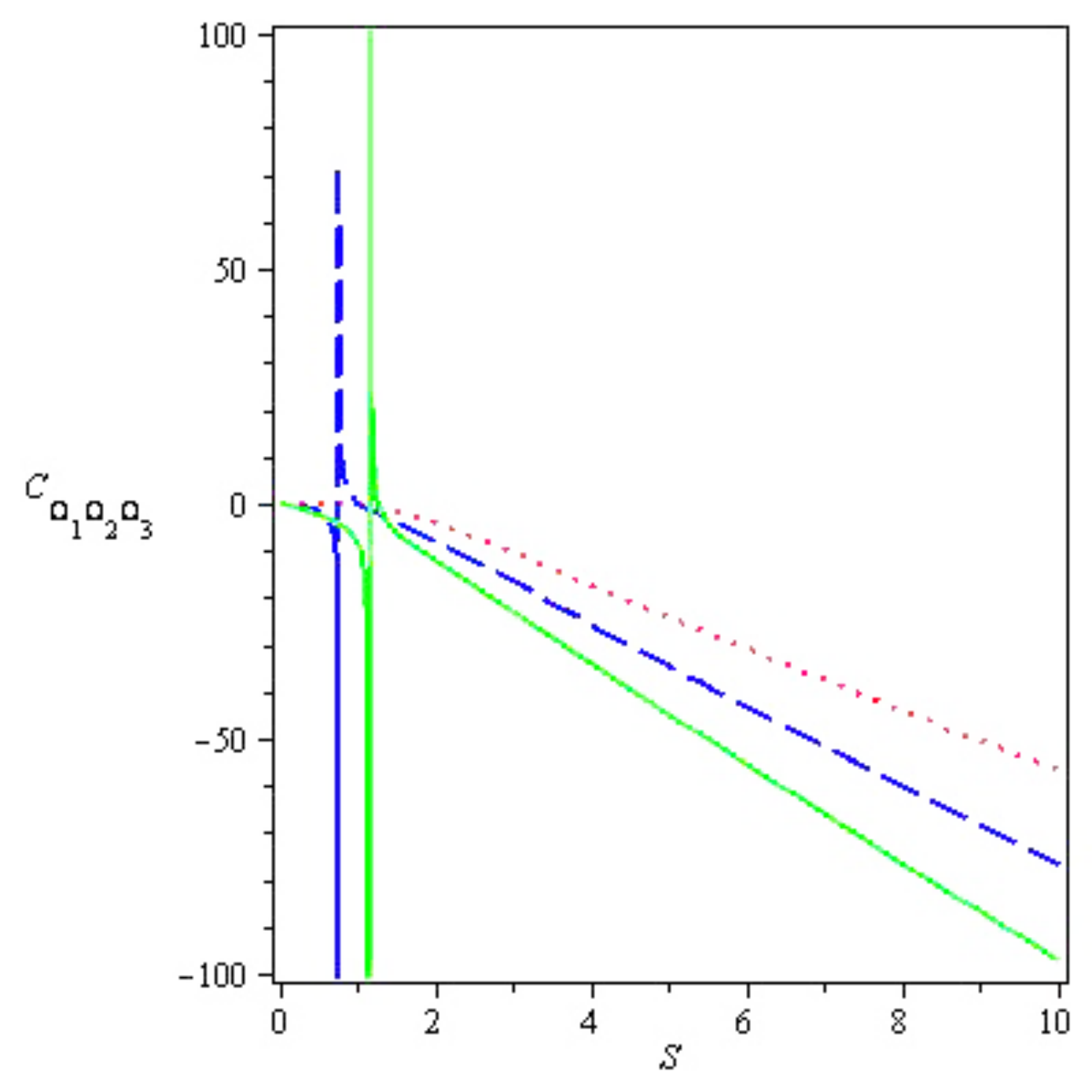} }
\caption{Graph of the the scalar curvature $C_{\Omega_{1},\Omega_{2},\Omega_{3}}$ with respect to entropy, $S$, for  $J_{1} =J_{2} = J_{3} $= 1. The dot red, dot-dashed blue, and solid green curves correspond to $d=8$, $d=10$ and $d=12$, respectively. }
\end{figure}
\begin{figure}[tbp]
\centering
\fbox{\includegraphics[scale=.47]{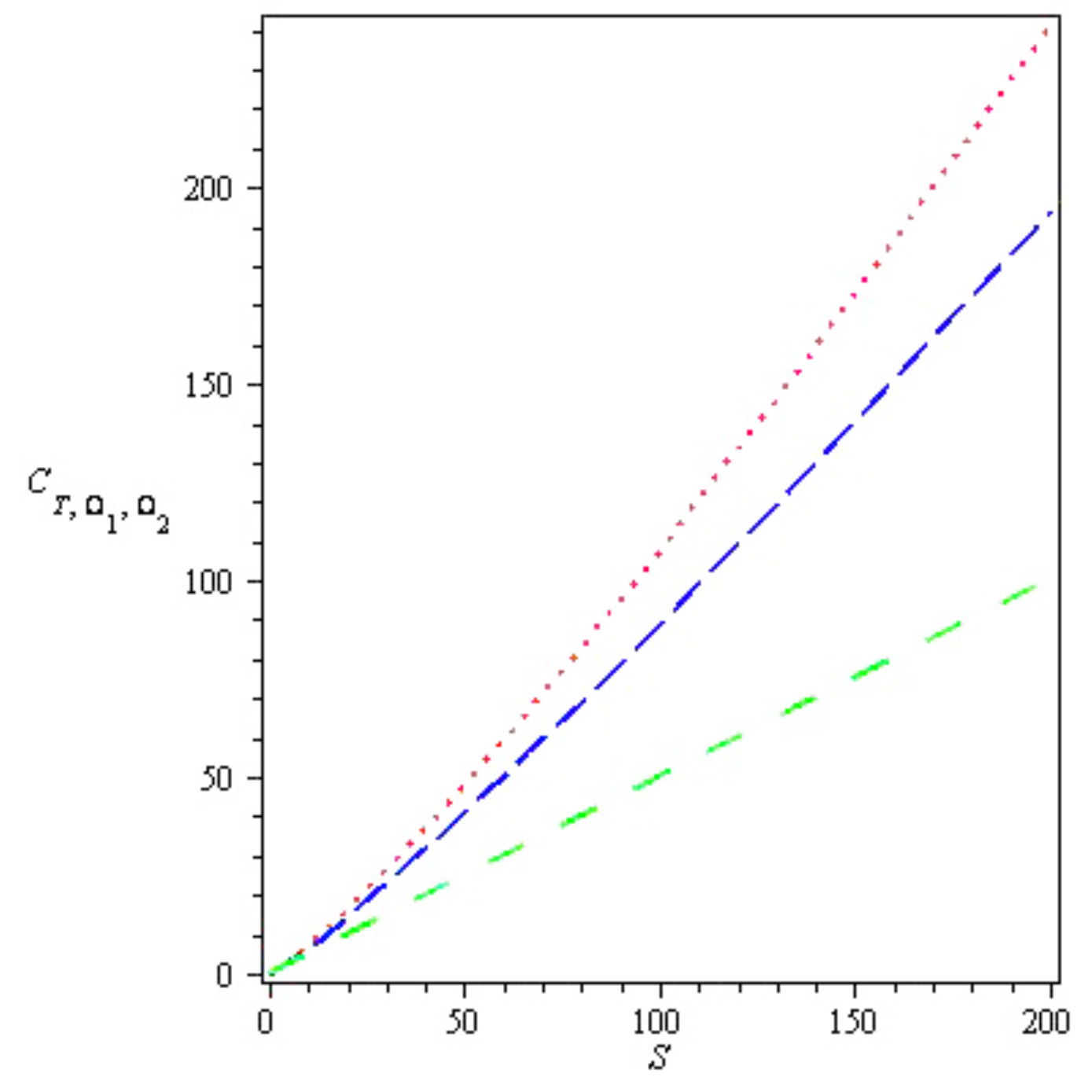}}
\caption{Graph of the the scalar curvature $C_{T,\Omega_{1},\Omega_{2}}$ with respect to entropy, $S$, for  $J_{1} =J_{2} $= 1, and $J_{3} $= 0. The dot red, dot-dashed blue, and solid green curves correspond to $d=8$, $d=10$ and $d=12$, respectively. }
\end{figure}
\begin{figure}[tbp]
\centering
\fbox{\includegraphics[scale=.47]{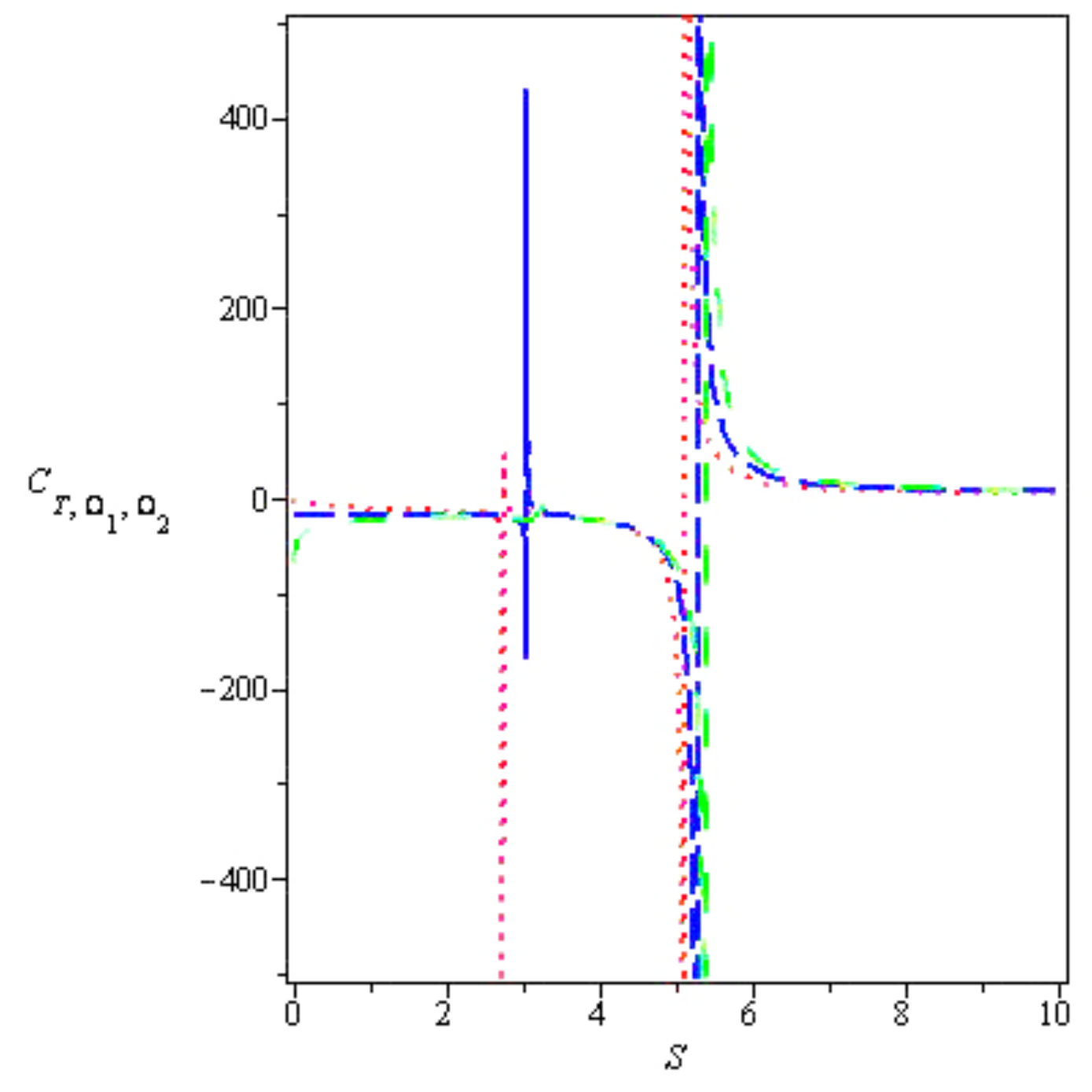}}
\caption{Graph of the the scalar curvature $C_{T,\Omega_{1},\Omega_{2}}$ with respect to entropy, $S$, for  $J_{1}$=1, $J_{2} $= 2, and $J_{3} $= 3. The dot red, dot-dashed blue, and solid green curves correspond to $d=7$, $d=9$ and $d=11$, respectively. }
\end{figure}
\begin{figure}[tbp]
\centering
\fbox{\includegraphics[scale=.47]{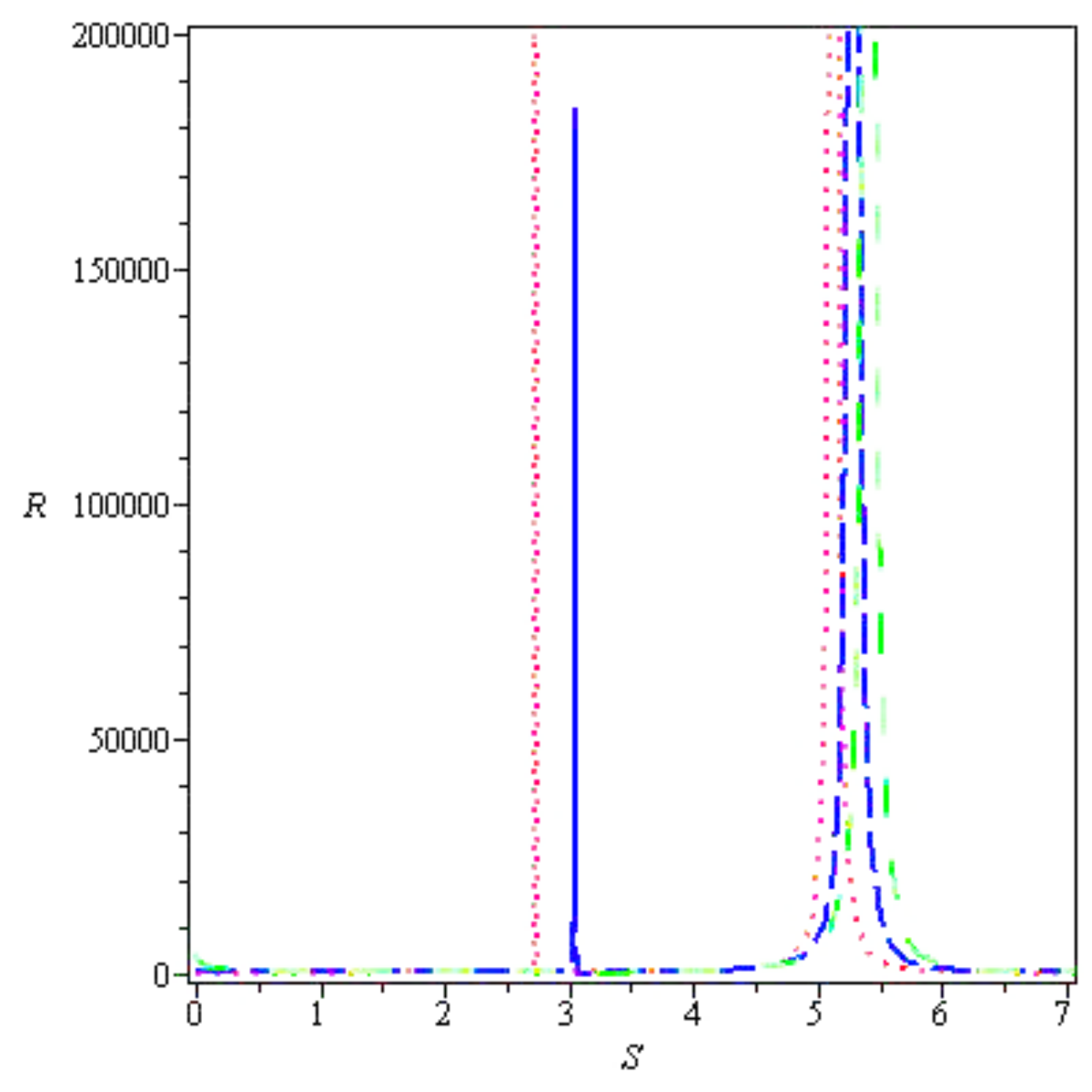} }
\caption{Graph of the the scalar curvature $R^{Q}(S,J_{1},J_{2},J_{3})$ with respect to entropy, $S$, for  $J_{1}$=1,$J_{2} $= 2, and $J_{3} $= 3. The dot red, dot-dashed blue, and solid green curves correspond to $d=7$, $d=9$ and $d=11$, respectively. }
\end{figure}

\section{Conclusion}
In this work, the Nambu brackets were exploited to obtain a simple representation of  the conformal transformations that connect different thermodynamic metrics to each other. We explored  two new formulations for the thermodynamic geometry of black holes in the electric charge and spin  representations. These formulations yield the correct relation between heat capacities and curvature singularities. Using the Nambu bracket approach, we also obtained some interesting exact results that relates  the Hessian matrix in an arbitrary dimensions to specific heats. Thus, one can easily identify the singularities of the thermodynamic scalar curvature.  The relationship between singularities of the scalar curvature in various representations of thermodynamic geometry and phase transitions of different heat capacities were also studied. We also investigated the Myers-Perry black hole with three spins. It will be interesting to  explore the thermodynamics of the Meyers-Perry black holes with more than three spins. We hope to report on our progress in this area in the near future. Our exact results  will be useful for a clear understanding of different aspects of the thermodynamic geometry of physical systems.

$ \\\\\\\\\\\\\
\\\\\\\\\\\\\
\\\\\\\\\\\\\\\\\\\\\\\\\\\\\\\
\\\\\\\\\\\\\\\\\\
\\\\\\$

\textbf{\large Appendix A }\\
In the following, we provide the proofs for identity (\ref{y1}) and the Jacobian transformation (\ref{N17}). We could write the following equation:
\beq
df={{\left( \frac{\partial f}{\partial h} \right)}_{g}}dh+{{\left( \frac{\partial f}{\partial g} \right)}_{h}}dg
\eeq
if $f$ is a function of $g$ and $h$. By considering a new function $k$, we rewrite the above equation as:
\beq
{{\left( \frac{\partial f}{\partial h} \right)}_{g}}{{\left( \frac{\partial h}{\partial g} \right)}_{k}}={{\left( \frac{\partial f}{\partial g} \right)}_{k}}-{{\left( \frac{\partial f}{\partial g} \right)}_{h}}
\eeq
Taking into account (\ref{N15}), we have:
\begin{eqnarray}
\frac{{{\left\{ f,g \right\}}_{a,b}}}{{{\left\{ g,h \right\}}_{a,b}}}\frac{{{\left\{ h,k \right\}}_{a,b}}}{{{\left\{ g,k \right\}}_{a,b}}}=\hspace{1cm}\\ \no \frac{{{\left\{ f,h \right\}}_{a,b}}{{\left\{ g,k \right\}}_{a,b}}-{{\left\{ f,k \right\}}_{a,b}}{{\left\{ g,h \right\}}_{a,b}}}{{{\left\{ g,k \right\}}_{a,b}}{{\left\{ g,h \right\}}_{a,b}}}
\end{eqnarray}
At last, we get the identity (\ref{y1}):
   \begin{eqnarray}
  {{\left\{ f,g \right\}}_{a,b}}{{\left\{ h,k \right\}}_{a,b}}=\hspace{1cm}\\ \no {{\left\{ f,h \right\}}_{a,b}}{{\left\{ g,k \right\}}_{a,b}}-{{\left\{ f,k \right\}}_{a,b}}{{\left\{ g,h \right\}}_{a,b}}
  \end{eqnarray}
  The Jacobian transformation can be written in the bracket notation in the form below:
\begin{eqnarray}
\frac{\partial \left( f,g \right)}{\partial (h,k)}={{\left( \frac{\partial f}{\partial h} \right)}_{k}}{{\left( \frac{\partial g}{\partial k} \right)}_{h}}-{{\left( \frac{\partial f}{\partial k} \right)}_{h}}{{\left( \frac{\partial g}{\partial h} \right)}_{k}}=\hspace{.4cm}\no \\ \no
\left( \frac{{{\left\{ f,k \right\}}_{a,b}}}{{{\left\{ h,k \right\}}_{a,b}}} \right)\left( \frac{{{\left\{ g,h \right\}}_{a,b}}}{{{\left\{ k,h \right\}}_{a,b}}} \right)-\left( \frac{{{\left\{ f,h \right\}}_{a,b}}}{{{\left\{ k,h \right\}}_{a,b}}} \right)\left( \frac{{{\left\{ g,k \right\}}_{a,b}}}{{{\left\{ h,k \right\}}_{a,b}}} \right)\\  =\frac{{{\left\{ f,g \right\}}_{a,b}}{{\left\{ h,k \right\}}_{a,b}}}{{{\left( {{\left\{ h,k \right\}}_{a,b}} \right)}^{2}}}=\frac{{{\left\{ f,g \right\}}_{a,b}}}{{{\left\{ h,k \right\}}_{a,b}}} \hspace{1.8cm}
\end{eqnarray}
We can also obtain the following equation using matrix product and the chain rule:
\begin{eqnarray}
&\no {{\left\{ {{f}_{1}},{{f}_{2}},...,{{f}_{n}} \right\}}_{{{h}_{1}},{{h}_{2}},...,{{h}_{n}}}}{{\left\{ {{h}_{1}},{{h}_{2}},...,{{h}_{n}} \right\}}_{{{q}_{1}},{{q}_{2}},...,{{q}_{n}}}}\\
&={{\left\{ {{f}_{1}},{{f}_{2}},...,{{f}_{n}} \right\}}_{{{q}_{1}},{{q}_{2}},...,{{q}_{n}}}}
\end{eqnarray}
Considering $f_{2}=h_{2}, ..., f_{n}=h_{n}$, yields
\begin{equation}
{{\left( \frac{\partial {{f}_{1}}}{\partial {{h}_{1}}} \right)}_{{{f}_{2}},{{f}_{3}},...,{{f}_{n}}}}=\frac{{{\left\{ {{f}_{1}},{{f}_{2}},...,{{f}_{n}} \right\}}_{{{q}_{1}},{{q}_{2}},...,{{q}_{n}}}}}{{{\left\{ {{h}_{1}},{{f}_{2}},...,{{f}_{n}} \right\}}_{{{q}_{1}},{{q}_{2}},...,{{q}_{n}}}}}
\end{equation}

\textbf{\large Appendix B }\\
The Maxwell's equations can be recast in the brackets notation.

Based on the first law of thermodynamics, $dM=TdS+\Phi dQ$, one could get the Maxwell's relation as follows:
\beq
{{\left( \frac{\partial T}{\partial Q} \right)}_{S}}={{\left( \frac{\partial \Phi }{\partial S} \right)}_{Q}}
\eeq
which is equivalent to:
\beq
{{\{T,S\}}_{S,Q}}=-{{\{\Phi ,Q\}}_{S,Q}}
\eeq
One can also derive the Maxwell's relations for the KN black hole as follows:
\beq\label{hk1}
{{\left\{ S,\Phi ,Q \right\}}_{S,Q,J}}=-{{\left\{ S,\Omega ,J \right\}}_{S,Q,J}}
\eeq
\beq
{{\left\{ \Phi ,Q,J \right\}}_{S,Q,J}}=-{{\left\{ T,S,J \right\}}_{S,Q,J}}
\eeq
\beq
{{\left\{ \Omega ,J,Q \right\}}_{S,Q,J}}={{\left\{ T,S,Q \right\}}_{S,Q,J}}
\eeq
Furthermore, one could extract some more Maxwell's relations:
\beq
{{\left\{ T,\Phi ,Q \right\}}_{S,Q,J}}=-{{\left\{ T,\Omega ,J \right\}}_{S,Q,J}}
\eeq
\beq
{{\left\{ \Phi ,Q,\Omega  \right\}}_{S,Q,J}}=-{{\left\{ T,S,\Omega  \right\}}_{S,Q,J}}
\eeq
\beq
{{\left\{ \Omega ,J,\Phi  \right\}}_{S,Q,J}}={{\left\{ T,S,\Phi  \right\}}_{S,Q,J}}
\eeq
\textbf{\large Appendix C}\\
Using Equation (\ref{N15}), we can define the specific heats with a fixed charge, electrical potential, temperature, and entropy for the Reissnr-Nordstrom black hole, respectively.
\bea\label{h4}
{{C}_{Q}}=T{{\left( \frac{\partial S}{\partial T} \right)}_{Q}}=\frac{T{{\left\{ S,Q \right\}}_{S,Q}}}{{{\left\{ T,Q \right\}}_{S,Q}}}
\eea
\bea\label{h5}
{{C}_{\Phi }}=T{{\left( \frac{\partial S}{\partial T} \right)}_{\Phi }}=\frac{T{{\left\{ S,\Phi  \right\}}_{S,Q}}}{{{\left\{ T,\Phi  \right\}}_{S,Q}}}
\eea
\beq\label{h7}
{{C}_{T}}={{\left( \frac{\partial Q}{\partial \Phi } \right)}_{T}}=\frac{{{\left\{ Q,T \right\}}_{S,Q}}}{{{\left\{ \Phi ,T \right\}}_{S,Q}}}
\eeq
\beq\label{h6}
{{C}_{S}}={{\left( \frac{\partial Q}{\partial \Phi } \right)}_{S}}=\frac{{{\left\{ Q,S \right\}}_{S,Q}}}{{{\left\{ \Phi ,S \right\}}_{S,Q}}}=\frac{{{\left\{ Q,S \right\}}_{S,\Phi }}}{{{\left\{ \Phi ,S \right\}}_{S,\Phi }}}
\eeq
In the last case, Relation (\ref{N20}), has been used.
We can also obtain some well-known relations between different heat capacities as follows:
\beq\label{h11}
{{C}_{\Phi }}-{{C}_{Q}}=-\frac{TQ{{\alpha }^{2}}}{k}
\eeq
\beq\label{h12}
{{C}_{\Phi }}-{{C}_{Q}}=\frac{T{{\alpha }^{2}}{{C}_{T}}}{{{k}^{2}}}
\eeq
\beq\label{h13}
{{C}_{T}}=-Qk
\eeq
\beq\label{h8}
{{C}_{Q}}{{C}_{T}}{{\left( {{C}_{\Phi }}{{C}_{S}} \right)}^{-1}}=1
\eeq
where, $\alpha$ and $k$ are the electric potential coefficient of expansion and isothermal compressibility, respectively. Using the bracket representation, they can be rewritten as follows:
\begin{eqnarray}
\label{h9}
\alpha =\frac{1}{Q}{{\left( \frac{\partial Q}{\partial T} \right)}_{\Phi }}=\frac{1}{Q}\frac{{{\left\{ Q,\Phi  \right\}}_{S,Q}}}{{{\left\{ T,\Phi  \right\}}_{S,Q}}}\\ \no =\frac{1}{Q}\frac{{{\left\{ T,S \right\}}_{S,Q}}}{{{\left\{ T,\Phi  \right\}}_{S,Q}}}=\frac{1}{Q}{{\left( \frac{\partial S}{\partial \Phi } \right)}_{T}}
\end{eqnarray}
\begin{eqnarray}
\label{h10}
k=-\frac{1}{Q}{{\left( \frac{\partial Q}{\partial \Phi } \right)}_{T}}=-\frac{1}{Q}\frac{{{\left\{ Q,T \right\}}_{S,Q}}}{{{\left\{ \Phi ,T \right\}}_{S,Q}}}\\ \no =\alpha \frac{{{\left\{ Q,T \right\}}_{S,Q}}}{{{\left\{ Q,\Phi  \right\}}_{S,Q}}}=\alpha {{\left( \frac{\partial T}{\partial \Phi } \right)}_{Q}}\hspace{.5cm}
\end{eqnarray}
In which Maxwell equations have been used to get the first equation.
It is suitable to prove Relations (\ref{h11})  and (\ref{h8}) ,respectively, by:
\begin{widetext}
\bea
   {{C}_{\Phi }}-{{C}_{Q}}=T\left[ {{\left( \frac{\partial S}{\partial T} \right)}_{\Phi }}-{{\left( \frac{\partial S}{\partial T} \right)}_{Q}} \right]=T\left[ \frac{{{\left\{ S,\Phi  \right\}}_{S,Q}}}{{{\left\{ T,\Phi  \right\}}_{S,Q}}}-\frac{{{\left\{ S,Q \right\}}_{S,Q}}}{{{\left\{ T,Q \right\}}_{S,Q}}} \right]= \hspace{.3cm} \\ \no
  T\left[ \frac{{{\left\{ S,\Phi  \right\}}_{S,Q}}{{\left\{ T,Q \right\}}_{S,Q}}-{{\left\{ S,Q \right\}}_{S,Q}}{{\left\{ T,\Phi  \right\}}_{S,Q}}}{{{\left\{ T,\Phi  \right\}}_{S,Q}}{{\left\{ T,Q \right\}}_{S,Q}}} \right]=T\left[ \frac{{{\left\{ S,T \right\}}_{S,Q}}{{\left\{ \Phi ,Q \right\}}_{S,Q}}}{{{\left\{ T,\Phi  \right\}}_{S,Q}}{{\left\{ T,Q \right\}}_{S,Q}}} \right] \\ \no
  =T\left[ \frac{{{\left\{ \Phi ,Q \right\}}_{S,Q}}{{\left\{ \Phi ,Q \right\}}_{S,Q}}}{{{\left\{ T,\Phi  \right\}}_{S,Q}}{{\left\{ T,Q \right\}}_{S,Q}}}\frac{{{\left\{ T,\Phi  \right\}}_{S,Q}}}{{{\{T,\Phi \}}_{S,Q}}} \right]=-\frac{TQ{{\alpha }^{2}}}{k}\hspace{2cm}
\eea
\end{widetext}
\begin{eqnarray}
{{C}_{Q}}{{C}_{T}}{{\left( {{C}_{\Phi }}{{C}_{S}} \right)}^{-1}}=\hspace{1.2cm} \\ \no \left[ \left( T\frac{{{\{S,Q\}}_{S,Q}}}{{{\{T,Q\}}_{S,Q}}} \right)\left( \frac{{{\{Q,T\}}_{S,Q}}}{{{\{\Phi ,T\}}_{S,Q}}} \right) \right]
\times \hspace{.6cm} \\ \no {{\left[ \left( T\frac{{{\{S,\Phi \}}_{S,Q}}}{{{\{T,\Phi \}}_{S,Q}}} \right)\left( \frac{{{\{Q,S\}}_{S,Q}}}{{{\{\Phi ,S\}}_{S,Q}}} \right) \right]}^{-1}}=1
\end{eqnarray}
Moreover, one could exploit the important relations between the heat capacities with three parameters as follows:
\begin{widetext}
\bea
& {{C}_{J,Q}}{{C}_{J,T}}{{\left( {{C}_{J,S}}{{C}_{J,\Phi }} \right)}^{-1}}=1 \,\ ; \,\  {{C}_{Q,J}}{{C}_{T,Q}}{{\left( {{C}_{Q,S}}{{C}_{Q,\Omega }} \right)}^{-1}}=1 \,\ ; \,\  {{C}_{S,Q}}{{C}_{S,\Omega }}{{\left( {{C}_{S,J}}{{C}_{S,\Phi }} \right)}^{-1}}=1\\
& \nonumber {{C}_{\Phi ,J}}{{C}_{\Phi ,T}}{{\left( {{C}_{\Phi ,S}}{{C}_{\Phi ,\Omega }} \right)}^{-1}}=1 \,\ ; \,\ {{C}_{T,Q}}{{C}_{T,\Omega }}{{\left( {{C}_{T,\Phi }}{{C}_{T,J}} \right)}^{-1}}=1 \,\ ; \,\ {{C}_{\Omega ,Q}}{{C}_{\Omega ,T}}{{\left( {{C}_{\Omega ,S}}{{C}_{\Omega ,\Phi }} \right)}^{-1}}=1
\label{ap1}
\eea

\end{widetext}




\end{document}